\documentclass[11pt,draftclsnofoot,peerreviewca,letterpaper,onecolumn]{IEEEtran}

\usepackage{cite}
\usepackage{graphicx,color,epsfig,rotating}
\usepackage{amsfonts,amsmath,amssymb,bbm}
\usepackage{algorithm,algorithmic}
\usepackage{subfigure}
\usepackage{amsmath}
\usepackage{cite}
\usepackage{mdwtab}
\usepackage{subfigure}
\usepackage{placeins}
\usepackage{psfrag, graphicx}
\usepackage[latin1]{inputenc}
\usepackage{amssymb}

\long\def\comment#1{}

\newfont{\bbb}{msbm10 scaled 700}

\newfont{\bb}{msbm10 scaled 1100}

\newcommand{\PP}{\mbox{\bb P}}

\newcommand{\FF}{\mbox{\bb F}}

\newcommand{\EE}{\mbox{\bb E}}

\newcommand{\cv}{{\bf c}}

\newcommand{\wv}{{\bf w}}

\newcommand{\Gm}{{\bf G}}

\newcommand{\Ac}{{\cal A}}

\newcommand{\Dc}{{\cal D}}

\newcommand{\Fc}{{\cal F}}

\newcommand{\Kc}{{\cal K}}
\newcommand{\Lc}{{\cal L}}

\newcommand{\Uc}{{\cal U}}

\newcommand{\fsf}{{\sf f}}

\newcommand{\ssf}{{\sf s}}

\newcommand{\Asf}{{\sf A}}

\newcommand{\Ssf}{{\sf S}}
\newcommand{\Tsf}{{\sf T}}
\newcommand{\Usf}{{\sf U}}

\newcommand{\eqdef}{\stackrel{\Delta}{=}}

\newcommand{\be}{\begin{equation}}
\newcommand{\ee}{\end{equation}}
\newcommand{\bea}{\begin{eqnarray}}
\newcommand{\eea}{\end{eqnarray}}

\def\fsf{ {\sf f}}

\newtheorem{defn}{Definition}
\newtheorem{example}{Example}
\newtheorem{theorem}{Theorem}
\newtheorem{lemma}{Lemma}
\newtheorem{corollary}{Corollary}

\renewcommand{\baselinestretch}{1.3}

\begin{document}

\title{Fundamental Limits of Caching in Wireless D2D Networks}

\author{Mingyue Ji,~\IEEEmembership{Student Member,~IEEE}, 
Giuseppe Caire,~\IEEEmembership{Fellow,~IEEE}, \\
and Andreas F. Molisch,~\IEEEmembership{Fellow,~IEEE}
\thanks{The authors are with the Department of Electrical Engineering,
University of Southern California, Los Angeles, CA 90089, USA. (e-mail: \{mingyuej, caire, molisch\}@usc.edu)}
}

\maketitle

\begin{abstract}
We consider a wireless Device-to-Device (D2D) network where communication is restricted to be single-hop. 
Users make arbitrary requests from a finite library of files and have pre-cached information on their devices, 
subject to a per-node storage capacity constraint.  A similar problem has already been considered in an ``infrastructure'' setting, where 
all users receive a common multicast (coded) message from a single omniscient server (e.g., a base station having all the files in the library) 
through a shared bottleneck link.  In this work, we consider a D2D ``infrastructure-less'' version of the problem. 
We propose a caching strategy based on deterministic assignment of subpackets of the library files, 
and a coded delivery strategy where the users send linearly 
coded messages to each other in order to collectively satisfy their demands. 
We also consider a random caching strategy, which is more suitable to a fully decentralized 
implementation. Under certain conditions, both approaches can achieve the information theoretic outer 
bound within a constant  multiplicative factor.

In our previous work, we showed that a caching D2D wireless network with one-hop communication, 
random caching, and uncoded delivery (direct file transmissions), achieves the same throughput scaling law of the 
infrastructure-based coded multicasting scheme, in the regime of large number of users and  files in the library. 
This shows that the {\em spatial reuse gain}  of the  D2D network is order-equivalent  to the {\em coded multicasting gain}  
of single base station transmission. It is therefore natural to ask whether these two gains are cumulative, i.e., 
if a D2D network with both local communication (spatial reuse) and coded multicasting can provide an improved scaling law.
Somewhat counterintuitively, we show that these gains do not cumulate (in terms of throughput scaling law). 
This fact can be explained by noticing that the coded delivery scheme creates messages that are useful to multiple nodes, such that it
benefits from broadcasting to as many nodes as possible, while spatial reuse capitalizes on the fact that the 
communication is local, such that the same time slot can be re-used in space across the network. 
Unfortunately, these two issues are in contrast with each other. 
\end{abstract}


\newpage

\section{Introduction}
\label{section: intro}

Wireless traffic is dramatically increasing, under the constant pressure of killer apps such as on-demand (pre-stored)
video streaming \cite{cisco66}.  One of the most promising approaches for solving this problem is \emph{caching}, i.e., 
storing the content files in the users' devices and/or in dedicated helper nodes disseminated in the network coverage area \cite{6495773, golrezaei2012femtocaching, golrezaei2012device, ji2013optimal, ji2013wireless , gitzenis2012asymptotic}. 
Imagine a moderately dense urban area, such as a university campus, 
where  $n \approx 10000$ users distributed over a surface of $\approx 1$ km$^2$ stream movies 
from a library of  $m \approx 100$ files, such as the Netflix or Amazon Prime weekly top-of-the chart titles.
Capitalizing on the fact that  user demands are highly redundant, each user demand can be satisfied through 
local communication from a cache, without requiring a high-throughput backhaul to the core network. 
Such backhaul would constitute a major bottleneck, being too costly or, in the case of wireless helper 
nodes and user devices, by definition infeasible. 

In \cite{ji2013optimal, ji2013optimalJ} we studied a one-hop Device-to-Device (D2D) communication network with caching at the user nodes. 
The network is formed by $n$ user nodes, each of which stores $M$ files from a library of $m$ files. 
Under the simple protocol model of \cite{gupta2000capacity}, we showed that by using a well-designed random caching policy 
and interference-avoidance transmission with spatial reuse, such that links sufficiently separated in space can be simultaneously active, as 
$n, m \rightarrow \infty$ with $nM \gg m$ the throughput per user behaves as $\Theta\left(\frac{M}{m} \right)$ while the outage probability, 
i.e.,  the probability that a user request cannot be served, can be fixed to some small positive constant. 
Furthermore, this scaling is shown to be order-optimal under the considered network model.\footnote{We will use the following standard ``order'' notation: given two functions $f$ and $g$, we say that: 1)  $f(n) = O\left(g(n)\right)$ if there exists a constant $c$ and integer $N$ such that  $f(n)\leq cg(n)$ for $n>N$. 2) $f(n)=o\left(g(n)\right)$ if $\lim_{n \rightarrow \infty}\frac{f(n)}{g(n)} = 0$. 
3) $f(n) = \Omega\left(g(n)\right)$ if $g(n) = O\left(f(n)\right)$. 4) 
$f(n) = \omega\left(g(n)\right)$ if $g(n) = o\left(f(n)\right)$. 
5) $f(n) = \Theta\left(g(n)\right)$ if $f(n) = O\left(g(n)\right)$ and~$g(n) = O\left(f(n)\right)$.}

A different approach to caching is taken in \cite{maddah2012fundamental}, which considers 
a system with a single omniscient 
transmitter  (e.g., a cellular base station having all the files in the library) 
serving  $n$ receivers (users) through a common bottleneck link.
Instead of caching individual files, the users store carefully designed sets of packets from all files in the library. 
Such sets form the receivers side information, such that for any arbitrary set of user demands a common 
multicast {\em coded} message can be sent from the base station to all users in order to satisfy their demands. 
This multicast coded message is formed by a sequence of linear combinations of the 
file packets.\footnote{It is interesting to notice that, for any set of user demands, this system reduces to
a special instance of the general index coding problem \cite{bar2011index, el2010index, lubetzky2009nonlinear, blasiak2010index, chaudhry2011complementary, jafar2013topological, haviv2012linear, arbabjolfaei2013capacity, 6620404} for which coding based on clique covering \cite{bar2011index} is optimal 
within a bounded factor.}  Therefore, this scheme is referred to as ``coded multicasting'' in the following.

The scheme of \cite{maddah2012fundamental} satisfies any arbitrary set of user demands with a number of transmitted 
coded symbols equal to $n \left(1-\frac{M}{m}\right)\frac{1}{1+\frac{Mn}{m}}$ times the size of a single file (expressed in bits).
Approximate optimality within a constant factor is shown by developing a cut-set lower bound on the min-max number 
of transmissions.  Notice that for $nM \gg m$, the throughput scaling is again given by $\Theta\left (\frac{M}{m} \right )$. 

In the regime of fixed $M$ and large $m$, a  conventional system serving each user demand as an individual TCP/IP connection from a 
central server (e.g., a node of a  content distribution network \cite{nygren2010akamai} placed in the core network), 
as currently implemented today,  yields per-user throughput scaling $\Theta \left (\frac{1}{n} \right )$. This is because the downlink throughput of the common 
bottleneck link  (a constant) must be shared by $n$ simultaneous user demands, whose sum rate scales linearly with $n$ irrespectively of caching. 
Hence, it is apparent that a conventional system is not able to exploit the inherent ``content reuse'' in the system, i.e., the fact that 
a large number of users ask for a limited number of library files.  In contrast, both the caching approaches of \cite{ji2013optimal, ji2013optimalJ} and of  \cite{maddah2012fundamental} yield $\Theta \left (\frac{M}{m} \right )$, 
which is a much better scaling for $nM \gg m$, i.e.,  in the regime of highly redundant demands, for which caching 
is expected to be efficient.  Notably, the per-user throughput in both caching schemes 
scales linearly with the per-user cache memory size $M$, which is expected to grow with time
according to Moore's law of VLSI integration.  

The D2D approach of  \cite{ji2013optimal, ji2013optimalJ} capitalizes on
of the spatial reuse of D2D achieved by local communication, while the approach of \cite{maddah2012fundamental} 
exploits global communication in order to multicast the coded messages, simultaneously useful to a large number of users. 
A natural question at this point is whether any further gain can be obtained by {\em combining} spatial 
reuse and coded multicasting.

\subsection{Overview of the Main Results}
\label{section: overview}

Motivated by the above question, in this paper we consider the same model of D2D wireless networks 
as in \cite{ji2013optimal, ji2013optimalJ},  with a caching and delivery scheme inspired by \cite{maddah2012fundamental}, 
based on subpacketization in the caching phase and (inter-session) coding in the delivery phase.  
Our main contributions are as follows:  1) if each node in the network can reach in a single hop all other nodes in the network, the proposed scheme achieves almost the same throughput of \cite{maddah2012fundamental}, without the need of a central base station; 
2) if the transmission range of each node is limited, such that concurrent short range transmissions can co-exist in a spatial reuse scheme, then the throughput has the same scaling law (with possibly different leading term constants) of the reuse-only case \cite{ji2013optimal, ji2013optimalJ} or the coded-only case \cite{maddah2012fundamental}. This result holds even if one optimizes the transmission range and therefore the spatial reuse of the system. 
Counterintuitively, this means that it is not possible to cumulate the spatial reuse gain and the coded multicasting gain, 
and that these two albeit different type of gains are equivalent as far as the throughput scaling law is concerned. 
Beyond scaling laws, in order to establish the best combination of reuse and coded multicasting gains, trading off the rate achieved on each local link
(decreasing function of distance) with the number of users that can be reached by a coded multicast message (increasing function of distance), 
must be sought  in terms of the actual throughput in bit/s/Hz (i.e., in the coefficients of the dominant terms of the throughput scaling for large $n,m$ and finite $M$, and not just in the scaling law itself). 

We consider both deterministic caching and (decentralized) random caching (as done in \cite{maddah2013decentralized} for the single bottleneck link case). 
In both cases, we show that for most regimes of the system parameters (apart from the regime of very small caches, which is not really relevant for applications),  the throughput achieved with both the proposed deterministic and random caching schemes is optimal  within a constant factor. 

The paper is organized as follows. Section~\ref{sec: Network Model and Problem Formulation} presents the network model and the formal 
problem definition.  We illustrate all the main results based on the deterministic caching scheme and its implications in Section~\ref{sec: Main Results}. In Section~\ref{sec: A Decentralized Random Caching Construction}, we discuss the decentralized caching scheme and the corresponding 
coded delivery approach. Section~\ref{sec: Conclusions} contains our concluding remarks and the main proofs are given in 
Appendices in order to keep the flow of exposition.

\subsection{Remarks}
\label{section: remarks}

Before proceeding in the presentation of our results,  we would like to make a few remarks to clarify obvious 
questions and anticipate possible concerns that this line of work (see for example \cite{maddah2012fundamental, maddah2013decentralized, pedarsani2013online, niesen2013coded, golrezaei2012femtocaching, 6495773, gitzenis2012asymptotic, llorcatulino13, llorcatulino132, llorcatulino14, ji2013optimal, ji2013optimalJ, ji2013wireless, ji2013fundamental, ji2014order1, ji2014order2, sengupta2013fundamental, Karamchandani2014, Hachem2014}
) 
may raise. 

First, we would like to point out that in this paper we refer to ``coding'' in the sense 
of ``inter-session network coding'', i.e., when the codeword is a function of symbols (or, ``subpackets'') 
from different source messages (files in the library). 
Often, coding at the application layer \cite{shokrollahi2006raptor, luby2006raptor, aditya2011flexcast} 
or right on top of the transport layer \cite{sundararajan2009network} 
is used in an intra-session only mode, 
in order to send linear combinations of subpackets
from the {\em same} source message and cope with packet losses 
in the network, but without mixing subpackets of different messages.  
We point out that this intra-session ``packet erasure coding'' 
has conceptually little to do with ``network coding'', although
it has been sometimes referred to as ``random linear network coding''  when the linear combinations are generated with randomly drawn  
coefficients over some finite field. 
In line with  \cite{maddah2012fundamental} and with the protocol model of our previous work \cite{ji2013optimal, ji2013optimalJ}, 
also in this paper any transmission within the appropriate range is assumed to be ``noiseless'' (i.e., perfectly decoded) 
and therefore we will not consider  packet erasure coding against channel impairments. 

Then, it is important to notice that this work, as well as \cite{ji2013optimal, ji2013optimalJ,maddah2012fundamental}, is based on an underlying 
time-scale decomposition for which the caching phase (i.e., placing information in the caches) is done ``a priori'', at some time scale much slower 
than the delivery phase (i.e., satisfying the users demands). For example, we may imagine that the caches content is updated every day, 
through a conventional cellular network used during off-peak time, such that the content library is refreshed by inserting new titles and deleting old ones. 
This scenario differs significantly with respect to the conventional and widely studied ``on-line'' caching policies, 
where the  cache content is updated along with the delivery process \cite{meyerson2001web,  dowdy1982comparative, dan1996dynamic, korupolu1999placement, baev2008approximation, borst2010distributed, almeroth1996use}.

Finally, we would like to mention here that the considerations made in  \cite{ji2013optimal,maddah2012fundamental,maddah2013decentralized} about
handling asynchronous demands holds verbatim in this paper, and shall not be repeated for the sake of brevity. 
It should be clear that although we consider (for simplicity of exposition) files of the same length, 
the schemes described in this paper generalize immediately (with the same fundamental performance) to the case of 
unequal length files and asynchronous demands.   

\section{Network Model and Problem Definition}
\label{sec: Network Model and Problem Formulation}

\begin{figure}
\centering
\subfigure[]{
\centering \includegraphics[width=7cm]{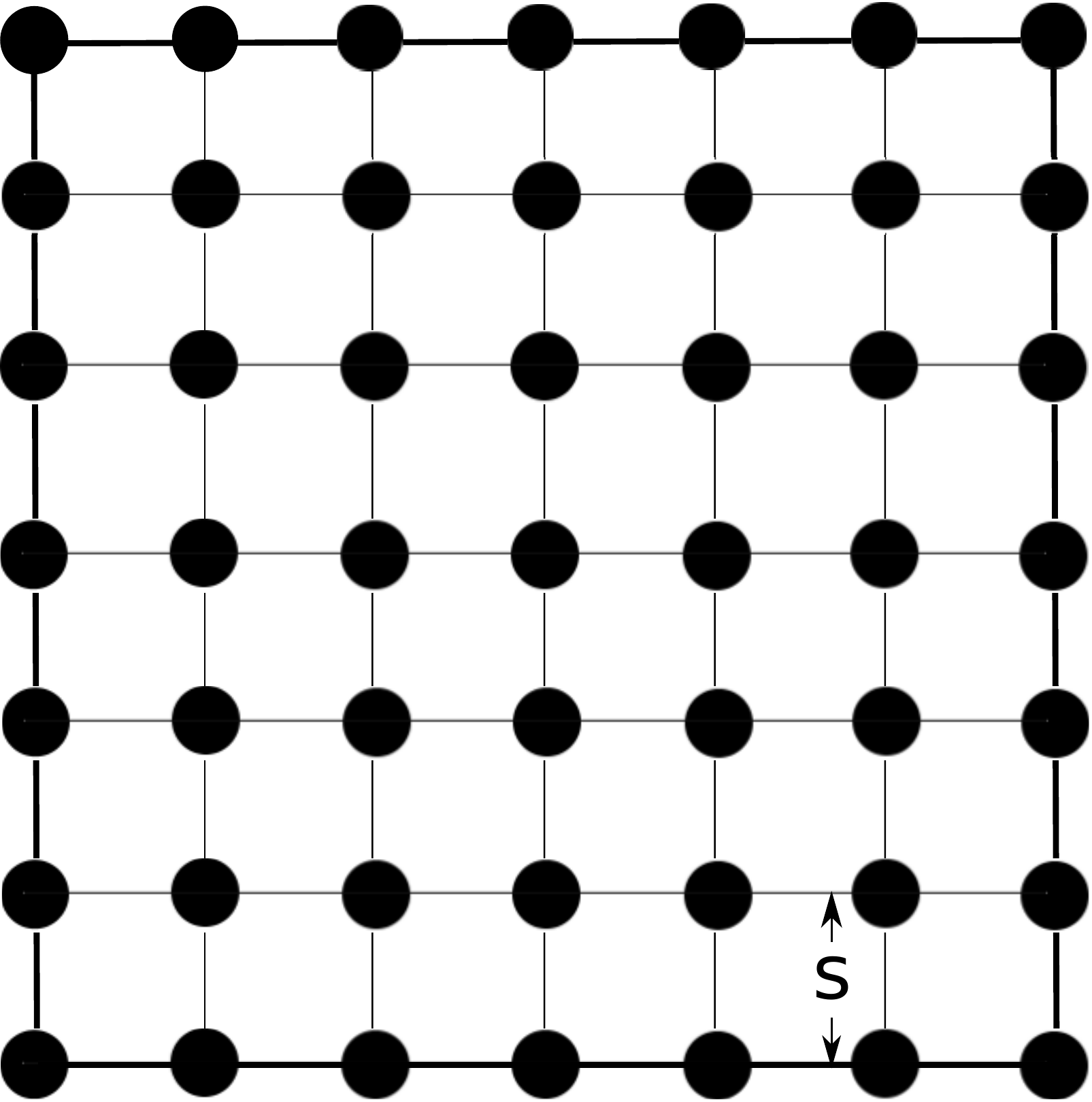}
\label{fig: Grid_Network_D2D}
}
\subfigure[]{
\centering \includegraphics[width=7cm]{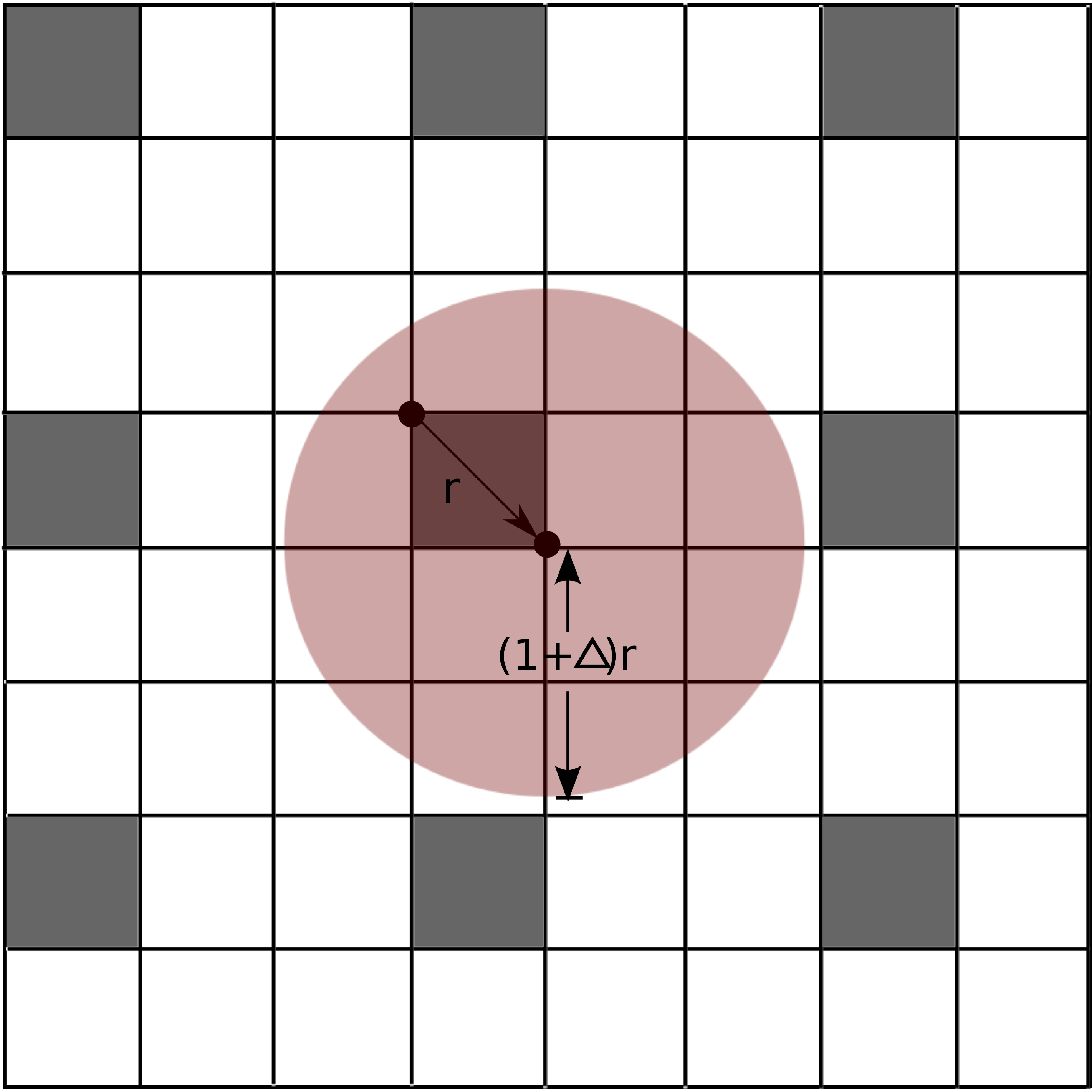}
\label{fig: Grid_TDMA}
}
\caption{a)~Grid network with $n=49$ nodes (black circles) with minimum separation $s = \frac{1}{\sqrt{n}}$. 
b)~An example of single-cell layout and the interference avoidance spatial reuse scheme. 
In this figure, each square represents a cluster. 
The gray squares represent the concurrent transmitting clusters. 
The red area is the disk where the protocol model allows no other concurrent transmission. 
$r$ is the worst case transmission range and $\Delta$ is the interference parameter. 
We assume a common $r$ for all the transmitter-receiver pairs. 
In this particular example, the reuse factor is $\Kc=9$.}
\end{figure}

We consider a grid network formed by $n$ nodes $\Uc = \{1, \ldots, n\}$ 
placed on a regular grid on the unit square, with minimum distance  $1/\sqrt{n}$. (see Fig.~\ref{fig: Grid_Network_D2D}).
Users $u \in \Uc$ make arbitrary requests $f_u \in \Fc = \{1, \ldots, m\}$, from a fixed file library of size $m$. 
The vector of requests is denoted by $\fsf = (f_{1}, \ldots, f_{n})$. 
Communication between user nodes obeys the following {\em protocol model}: if a node $i$ transmits a packet to node $j$, 
then the transmission is successful if and only if: a) The distance between $i$ and $j$ is less than $r$; b) Any other node 
$k$ transmitting simultaneously, is at distance $d(k,j) \geq (1+\Delta) r$ from the receiver $j$, where 
$r, \Delta > 0$ are protocol parameters. In practice, nodes send data at some constant rate $C_r$ bit/s/Hz, where $C_r$ is a non-increasing function of the transmission range $r$.

Unlike live streaming,  in video on-demand, the probability that two users wish to stream simultaneously a file at the {\em same time} 
is essentially zero, although there  is a large  redundancy in the demands when $n \gg m$. 
We refer to this feature of video on-demand streaming as 
the {\em asynchronous content reuse}.  In order to model the asynchronous content reuse built into the problem, and 
forbid any form of ``naive multicasting'', i.e., achieving {\em uncoded multicasting gain} by overhearing ``for free''  transmissions 
dedicated to other users,  we assume the following streaming model: 
1) Each file in the library is formed by $L$ packets;\footnote{This is compliant with current video streaming protocols such as 
DASH  \cite{6495773}, where the video file is split into segments which are sequentially downloaded by the streaming users.} 
2) Each user downloads an arbitrarily selected segment of length $L'$ packets of the requested file;
3) We shall consider the system performance in the case of large $L$ and arbitrary, but finite, $L'$. 
In addition, we consider the worst-case system
throughput over the users' demands. Hence, for sufficiently large $L$ and finite $L'$ and $m$ it is always 
possible to have non-overlapping segments, even though users may request the same file index $f$. 

As a consequence of the above model, a demand vector $\fsf$ is associated with a list of pointers  $\ssf$ with 
elements $s_u \in \{1, \ldots, L - L'+1\}$ such that,  for each $u$,  the demand to file $f_u$ implies that the packets 
$s_u, s_u+1, \ldots, s_u+L'-1$ of file $f_u$ are sequentially 
requested by user $u$.  For simplicity, the explicit dependency on $\ssf$ is omitted whenever there is no ambiguity of notation. 
We let $W_f^j$ denote packet $j$ of file $f \in \Fc$. Without loss of generality, we assume that each packet contains 
$F$ information bits,  such that $\{W_f^j\}$ are i.i.d. random variables uniformly distributed over $\{1, 2, 3, \cdots, 2^F\}$. 
As said before, we are generally interested in the case of large $L$ and finite $L'$. 
We have:

\begin{defn}
\label{def: Caching Phase}
{\bf (Caching Phase)} The caching phase is a map of the file library 
$\Fc$ onto  the cache of the users in $\Uc$.  Each cache has size $M$ files. 
For each $u \in \Uc$, the function $\phi_u: \FF_2^{mFL} \rightarrow \FF_2^{MFL}$ 
generates the cache content $Z_u \triangleq \phi_u(W_f^j, f = 1, \cdots, m,  j = 1, \cdots, L)$. 
\hfill $\lozenge$
\end{defn}

\begin{defn}
\label{def: Delivery Phase}
{\bf (Coded Delivery Phase)} The delivery phase is defined by two sets of functions: 
the node encoding functions, denoted by $\{\psi_u: u \in \Uc\}$, and the node decoding functions, 
denoted by $\{\lambda_u: u \in \Uc\}$. 
Let $R_u^{\rm T}$ denote the number of {\em coded bits} 
transmitted by node $u$ to satisfy the request vector $\fsf$. 
The rate of node $u$ is defined by $R_u = \frac{R_u^{\rm T}}{FL'}$.  
The function $\psi_{u}: \FF_2^{MFL} \times \Fc^n \rightarrow \FF_2^{FL'R_u}$ generates the transmitted message 
$X_{u,\fsf}  \triangleq \psi_{u}(Z_u,\fsf)$ of node $u$ as a function of its cache content $Z_u$ and of the demand vector $\fsf$. 

Let $\Dc_u$ denote the set of users whose transmit messages are received by user $u$
(according to some transmission policy in Definition \ref{def:txpolicy}). 
The function $\lambda_u : \FF_2^{FL'\sum_{v \in \mathcal{D}_u} R_v} \times \FF_2^{MFL} \times \Fc^n \rightarrow \FF_2^{FL'}$
decodes the request of user $u$ from the received messages and its own cache, i.e., we have
\be
\hat{W}_{u,\fsf} \triangleq \lambda_u(\{ X_{v,\fsf} : v \in \Dc_u\}, Z_u, \fsf). 
\ee
\hfill $\lozenge$
\end{defn}
The worst-case error probability is defined as
\be
\label{eq: error prob}
P_e = \max_{\fsf \in \Fc^n,  \ssf \in \{1, \ldots, L - L'+1\}^n} \; \max_{u \in \Uc} \; \PP\left( \hat{W}_{u,\fsf}  \neq (W_{f_u}^{s_u}, \ldots, W_{f_u}^{s_u+L'-1}) \right).
\ee
Letting $R = \sum_{u \in \Uc} R_u$, the cache-rate pair $(M, R)$ is achievable if $\forall$ $\varepsilon > 0$ 
there exist a sequence indexed by the packet size $F \rightarrow \infty$ of cache encoding functions $\{\phi_u\}$, 
delivery functions $\{\psi_u\}$ and decoding functions $\{\lambda_u\}$, with rate $R^{(F)}$ and probability of error $P_e^{(F)}$ 
such that $\limsup_{F \rightarrow \infty} R^{(F)} \leq R$ and  $\limsup_{F \rightarrow \infty} P_e^{(F)} \leq \varepsilon$.
The optimal achievable rate~\footnote{As a matter of fact, this is the min-max number of packet transmissions where 
min is over the caching/delivery scheme and max is over the demand vectors, and thus intuitively is the inverse of the "rate" commonly used in communications theory. We use the term ``rate'' in order to stay compliant with the terminology introduced in 
\cite{maddah2012fundamental}.} is given by 
\be
R^*(M) \triangleq \inf\{R : (M, R) \text{ is achievable}\}.
\ee
In order to relate this definition of rate to the throughput of the network, defined later, 
we borrow from \cite{ji2013optimal, ji2013optimalJ} the definition of transmission policy: 

\begin{defn} \label{def:txpolicy}
{\bf (Transmission policy)} The transmission policy $\Pi_t$ is a rule to activate 
the D2D links in the network. Let $\Lc$ denote the set of all directed links.  
Let $\Ac \subseteq 2^\Lc$ the set of all possible feasible subsets of links 
(this is a subset of the power set of $\Lc$, formed by all sets of links forming independent sets in the 
network interference graph induced by the protocol model).  
Let $\Asf \subset \Ac$ denote a feasible set of simultaneously active links. 
Then, $\Pi_t$ is a conditional probability mass function over $\Ac$ given $\fsf$ (requests) and the caching functions,
assigning probability $\Pi_t(\Asf)$ to $\Asf \in \Ac$. 
\hfill $\lozenge$
\end{defn}

All the achievability results of this work are obtained using deterministic transmission policies, 
which are obviously a special case of Definition \ref{def:txpolicy}. 
Suppose that $(M,R)$ is achievable with a particular caching and delivering scheme. 
Suppose also that for a given transmission policy $\Pi_t$, the $R FL'$ coded bits to satisfy the worst-case 
demand vector can be delivered in $t_s$  
channel uses (i.e., it takes collectively $t_s$ channel uses in order to deliver the required $FL' R_u$ coded bits
to each user $u \in \Uc$, where each channel use carries $C_r$ bits). 
Then, the throughput per user, measured in useful information bits per channel use, 
is given by  
\begin{align}  
\label{useful-throughput-i}
T &\triangleq \frac{FL'}{t_s}. 
\end{align}
The pair $(M, T)$  is achievable if  $(M,R)$ is achievable and if there exists a transmission policy $\Pi_t$ such that the $RFL'$ encoded bits 
can be delivered to their destinations in $t_s \leq (F L')/T$ channel uses.  Then, the optimal achievable throughput is defined as
\be
T^*(M) \triangleq \sup\{T : (M, T) \text{ is achievable}\}
\ee

In the following we assume that $t = \eqdef \frac{Mn}{m} \geq 1$. 
Notice that this is a necessary condition in order to satisfy any arbitrary demand vector. 
In fact, if $t < 1$, then the aggregate cache  in the entire network cannot cache the file library, 
such that some files or part of files are missing and cannot be delivered. 
This requirement is not needed when there is an omniscient node that can supply the missing bits (as in \cite{maddah2012fundamental})
or in the case of random demands, as in \cite{ji2013optimal, ji2013optimalJ}, by defining a throughput versus outage probability tradeoff, 
where the outage probability is defined as the probability that a user demand cannot be satisfied. 
However, this work focuses on deterministic (worst-case) demands and has no omniscient node, such that $t \geq 1$ 
is necessary. 

We observe that our problem includes two parts: 1) the design of the caching, delivery and decoding functions; 
2) scheduling concurrent transmissions in the D2D network. 
For simplicity, we start by focusing on the case where only a single link can be simultaneously active in the whole network and let the transmission 
range $r$ such that any node can be heard by all other nodes (i.e., we let $r \geq \sqrt{2}$). In this case, scheduling of concurrent transmissions
in the D2D network is irrelevant and we shall focus only on the caching and delivery schemes. 
Then, we will relax the constraint on the transmission range  $r$ and consider spatial reuse 
and D2D link scheduling.

\section{Deterministic Caching, Achievability and Converse Bound}
\label{sec: Main Results}

\subsection{Transmission range $r \geq \sqrt{2}$}

The following theorem yields the achievable rate of the proposed 
caching and coded multicasting delivery scheme.
\begin{theorem}
\label{theorem: 1}
For $r \geq \sqrt{2}$ and $t = \frac{Mn}{m} \in \mathbb{Z}^+$, the following rate is achievable:
\be
\label{eq: theorem 1}
R(M) = \frac{m}{M}\left(1-\frac{M}{m}\right).
\ee
Moreover, when $t$ is not an integer, the convex lower envelope of $R(M)$, seen as a function of $M \in [0:m]$, is achievable.
\hfill  $\square$
\end{theorem}

The caching and delivery scheme achieving (\ref{eq: theorem 1}) is given in Appendix \ref{Caching Placement and Delivery Scheme} 
and an illustrative example is given in Section \ref{sec:example}. The proof of Theorem \ref{theorem: 1} is given in Appendix \ref{sec: Proof of Theorem 1 and Corollary 1}.
The corresponding achievable throughput is given by the following immediate corollary:
\begin{corollary}
\label{corollary: 1}
For $r \geq \sqrt{2}$,  the throughput 
\be  \label{cor 1}
T(M) = \frac{C_r}{R(M)},
\ee
where $R(M)$ is given by (\ref{eq: theorem 1}) is achievable.
\hfill  $\square$
\end{corollary}
\begin{IEEEproof} In order to deliver $FL'R(M)$ coded bits without reuse (at most one active link transmitting at any time)
we need $t_s = FL'R(M)/C_r$ channel uses. Therefore, (\ref{cor 1}) follows from the definition (\ref{useful-throughput-i}). 
\end{IEEEproof}
A lower bound (converse result) for the achievable rate in this case is given by the following theorem:
\begin{theorem}
\label{theorem: 2}
For $r \geq \sqrt{2}$, the optimal rate is lower bounded by 
\begin{align}
R^*(M) \geq & \max\left\{\max_{l \in \{1, 2, \cdots, \min\{m, n\}\}} \left(l - \frac{l}{\lfloor\frac{m}{l}\rfloor}M\right), 
\frac{n}{n-1}\left(1-\frac{M}{m}\right) \times 1\{n>1, m > 1\}\right\},  \label{banana}
\end{align}
where $1\{\cdot\}$ denotes an indicator function.
\hfill  $\square$
\end{theorem}
The proof of Theorem \ref{theorem: 2} is given in  Appendix \ref{Proof of Theorem 2}.
Given the fact that activating a single link per channel use is the best possible feasible
transmission policy, we obtain trivially that using the lower bound (\ref{banana}) {\em in lieu} of $R(M)$ in 
(\ref{cor 1}) we obtain an upper bound to any achievable throughput. 
The order optimality of our achievable rate is shown by:
\begin{theorem}
\label{theorem: gap 1}
As $n,m \rightarrow \infty$, for $t = \frac{Mn}{m} \geq 1$, the ratio of the achievable over the optimal rate is upper bounded by 
\be
\label{eq: gap 41}
\frac{R(M)}{R^*(M)} \leq \left\{\begin{array}{cc}4, &   t = \omega(1), \frac{1}{2} \leq M = o(m) \\ 
\frac{4t}{\lfloor t \rfloor}, & n = O(m), t = \Theta(1), \frac{1}{2} \leq M = o(m) \\
6, & M = \Theta(m) \\
\frac{2}{M}, & n = \omega(m), M <\frac{1}{2} \\
\frac{t}{\lfloor t \rfloor} \frac{2}{M}, &n = O(m), n > m, M <\frac{1}{2}\\
2, & n = O(m), n \leq m, M < \frac{1}{2} 
\end{array}\right.,
\ee
where, for all $t \geq 1$, we have $\frac{t}{\lfloor t \rfloor} \leq 2$. 
\hfill  $\square$
\end{theorem}
The proof of Theorem \ref{theorem: gap 1} is given in 
Appendix \ref{sec: proof of theorem 3}. Obviously, the same quantity upper bounds the optimal/achievable 
throughput ratio $\frac{T^*(M)}{T(M)}$. 

From (\ref{eq: gap 41}), we can see that except the case when $n > m$ and the storage capacity is very small ($M < \frac{1}{2}$, less than a half of a file), 
our achievable result can achieve the lower bound within a constant factor. 
The reason why in the regime of redundant requests ($n > m$) and small caches ($M < \frac{1}{2}$), 
the (multiplicative) gap is not bounded by a constant is because in our problem definition we force asynchronous requests (i.e., we let $L \rightarrow \infty$ with finite $L'$).
This prevents the possibility of ``naive multicasting'', i.e., sending the $n$ files directly, such that each file transmission is useful to multiple users that
requested that particular file.  This fact is evidenced by considering the special case of $L'=L$.  
In this case, naive multicasting becomes a valid scheme and we have:~\footnote{All the definitions in 
Section \ref{sec: Network Model and Problem Formulation} will be changed accordingly to the case when $L'=L$.}
\begin{corollary}
\label{corollary: naive multicast achievable}
For $r \geq \sqrt{2}$ and $t = \frac{Mn}{m} \in \mathbb{Z}^+$, the following rate is achievable:
{
\be
\label{eq: naive multicast achievable}
R(M) = \min\left\{ \frac{m}{M}\left(1-\frac{M}{m}\right), m  \right\}.
\ee
}
Moreover, when $t$ is not an integer, the convex lower envelope of $R(M)$, seen as a function of $M \in [0:m]$, is achievable.
\hfill  $\square$
\end{corollary}
{
The first term in the minimum in (\ref{eq: naive multicast achievable}) follows from Theorem \ref{theorem: 1}, while the second 
term is the rate obtained by using naive multicasting, where all the bits of all the files in the library are multicasted to all nodes, thus automatically satisfying any 
arbitrary request. This requires a total length of  $m LF$, i.e., a rate equal to $m$.}
Putting together Corollary \ref{corollary: naive multicast achievable} and Theorem \ref{theorem: 2} we have:

\begin{corollary}
\label{corollary: gap naive multicast}
For $r \geq \sqrt{2}$ and $t = \frac{Mn}{m} \geq 1$, as $n,m \rightarrow \infty$, the ratio of the achievable over the optimal rate is upper bounded by 
\be
\label{eq: gap 3}
\frac{R(M)}{R^*(M)} \leq \left\{\begin{array}{cc}4, &  t = \omega(1), \frac{1}{2} \leq M = o(m) \\
\frac{4t}{\lfloor t \rfloor}, & n = O(m), t = \Theta(1), \frac{1}{2} \leq M = o(m) \\
6, & M = \Theta(m) \\
2, & M <\frac{1}{2}
\end{array}\right.,
\ee
where $\frac{t}{\lfloor t \rfloor} \leq 2$. 
\hfill  $\square$
\end{corollary}

Corollary \ref{corollary: gap naive multicast} is also proved in Appendix  \ref{sec: proof of theorem 3}. 
Corollary \ref{corollary: gap naive multicast} implies that, 
when all the users request a whole file ($L = L'$), 
our achievable rate achieves the lower bound with a constant 
multiplicative factor in all the regimes of the system parameters. 

Beyond the theoretical interest of characterizing the system throughput in all regimes, we would like to remark here that,
in practice, caching is effective in the regime of large asynchronous content reuse (i.e., 
$n > m$) and moderate to large cache capacity (i.e., $1 \ll M < m$). In this relevant regime, we can focus on the 
asynchronous content delivery (no naive multicasting) letting $L \rightarrow \infty$ and fixed $L'$, and still
obtain a constant multiplicative gap from optimal.

\subsection{Transmission range $r < \sqrt{2}$}

In this case, the transmission range can be chosen in order to have localized  D2D communication and therefore 
allow for some spatial reuse. In this case, we need to design also a transmission policy to schedule concurrent active links. 
The proposed policy is based on {\em clustering}: the network is divided into clusters of equal size $g_c$, 
independently of the users' demands.  Users can receive messages only from nodes in the same cluster. 
Therefore, each cluster is treated as a small network. 
Assuming that $g_c M \geq m$,\footnote{If the condition $g_c M \geq m$ is not satisfied, 
we can choose a larger transmission range such that this condition is feasible.}
the total cache capacity of each cluster is sufficient to store the whole file library. 
Under this assumption, the same caching and delivery scheme used to prove Theorem \ref{theorem: 1} 
can be used here.  A simple achievable transmission policy consists of partitioning the set of clusters into
$\Kc$ reuse sets, such that the clusters of the same reuse set do not interfere and can be active simultaneously. In each active cluster, 
a single transmitter is active per time-slot and it is received by all the nodes in the cluster, as in classical  time-frequency reuse 
schemes with reuse factor $\Kc$ currently used in cellular networks \cite[Ch. 17]{molisch2011wireless}. 
An example of a reuse set is shown in Fig.~\ref{fig: Grid_TDMA}. In particular, we can pick $\Kc = \left(\left\lceil\sqrt{2}(1+\Delta)\right\rceil+1\right)^2$. 
This scheme achieves the following throughput:

\begin{theorem}
\label{theorem: 3}
Let $r$ such that any two nodes in a ``squarelet'' cluster of size $g_c$ can communicate, and  
let $t = \frac{g_cM}{m} \in \mathbb{Z}^+$. Then, the throughput
\be  \label{suca-throughput}
T(M) = \frac{C_r}{\Kc} \frac{1}{R(M)},
\ee
is achievable,  where $R(M)$ is given by {Theorem}~\ref{theorem: 1}, 
$r$ is the transmission range and $\Kc$ is the reuse factor. 
Moreover, when $t \notin \mathbb{Z}^+$, $T(M)$ is given by the expression (\ref{suca-throughput}) 
with $R(M)$ replaced by its lower convex envelope over $M \in [0:m]$.
\hfill  $\square$
\end{theorem}
The proof of Theorem \ref{theorem: 3} is given in Appendix \ref{sec: Proof of Theorem 4}.
Notice that whether reuse is convenient or not in this context depends on whether 
$C_{\sqrt{2}}$ (the link spectral efficiency for communicating across the network) is larger or smaller than $C_r/\Kc$, for some smaller $r$ which determines the 
cluster size. In turns, this depends on the how the link spectral efficiency varies as a function of the communication range. 
This aspect is not captured by the protocol model,  and the answer may depend on the operating frequency and appropriate channel 
model of the underlying wireless network physical layer \cite{ji2013wireless}. 

An upper bound on the throughput with reuse is given by:
\begin{theorem}
\label{theorem: 4}
When $r < \sqrt{2}$ and the whole library is cached within radius $r$ of any node, 
the optimal throughput is upper bounded by 
\be
T^*(M) \leq \frac{C_r \left\lceil \frac{4(2+\Delta)^2}{\Delta^2} \right\rceil}{\max_{l \in \{1, 2, \cdots, \min\{m, \lceil \pi r^2 n\rceil \}\}} \left(l - \frac{l}{\lfloor\frac{m}{l}\rfloor}M\right)},
\ee
where $r$ is the transmission range and $\Delta$ is the interference parameter. 
\hfill  $\square$
\end{theorem}
The proof of Theorem \ref{theorem: 4} is given in Appendix \ref{sec: Proof of theorem 5}. Furthermore, we have:
\begin{theorem}
\label{theorem: gap 2}
When $r < \sqrt{2}$, for $t = \frac{M \pi r^2 n }{m} \geq 1$, as $n,m \rightarrow \infty$, the ratio of the optimal throughput over the achievable throughput  is upper bounded by
\be
\frac{T^*(M)}{T(M)} \leq \Kc \left\lceil \frac{4(2+\Delta)^2}{\Delta^2} \right\rceil \times \left\{\begin{array}{cc}4, &   t = \omega(1), \frac{1}{2} \leq M = o(m) \\ 
\frac{4t}{\lfloor t \rfloor}, &  \pi r^2 n  = O(m), t = \Theta(1), \frac{1}{2} \leq M = o(m) \\
6, & M = \Theta(m) \\
\frac{2}{M}, &  \pi r^2 n  = \omega(m), M <\frac{1}{2} \\
\frac{t}{\lfloor t \rfloor} \frac{2}{M}, &n = O(m),  \pi r^2 n  > m, M <\frac{1}{2}\\
2, & n = O(m),  \pi r^2 n  \leq m, M < \frac{1}{2} 
\end{array}\right.,
\ee
where, for all $t \geq 1$, $\frac{t}{\lfloor t \rfloor} \leq 2$. 
\hfill  $\square$
\end{theorem}
The proof of Theorem \ref{theorem: gap 2} is given in Appendix \ref{sec: Proof of theorem gap}. Similar to the case of $r \geq \sqrt{2}$, 
when $L=L'$ (i.e., when naive multicasting is possible), we can  show that $\frac{T^*(M)}{T(M)}$ is upper bounded by 
the constant factor, 
independent of $m$, $n$ and $M$.

\subsection{An Example}  \label{sec:example}

\begin{figure}[ht]
\centerline{\includegraphics[width=10cm]{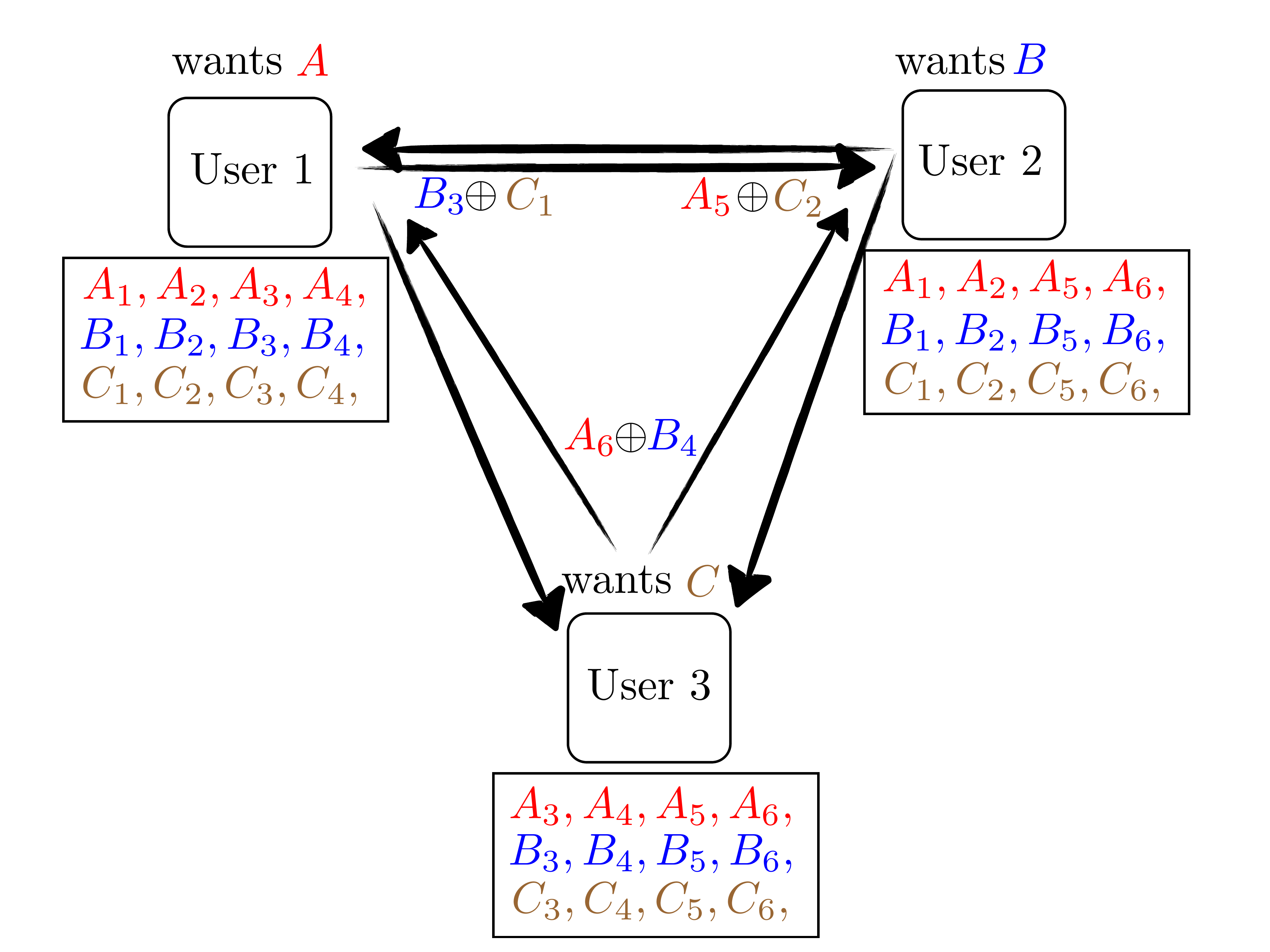}}
\caption{Illustration of the example of three users with $M = 2$, achieving rate $R(2) = 1/2$. }
\label{abc-user}
\end{figure}

The proposed caching placement and delivery scheme and the techniques of the proof for the converse are illustrated through a simple example.
Consider a network with three users ($n=3$). 
Each user can store $M=2$ files, and the library has size $m=3$ files, 
which are denoted by $A, B, C$.  Let $r \geq \sqrt{2}$.
Without loss of generality, we assume that each node requests one packet of a file ($L' = 1$).
We divide each packet of each file into $6$ subpackets, and denote the subpackets of the $j$-th 
packet as 
$\{A_{j,\ell} : \ell = 1, \ldots, 6\}$,
$\{B_{j,\ell} : \ell = 1, \ldots, 6\}$, and
$\{C_{j,\ell} : \ell = 1, \ldots, 6\}$.
The size of each subpacket is $F/6$. 
We let user $u$ stores $Z_u$, $u = 1, 2, 3$, given as follows:
\begin{align}
Z_1 = &(A_{j,1}, A_{j,2}, A_{j,3}, A_{j,4}, B_{j,1}, B_{j,2}, B_{j,3}, B_{j,4}, \notag\\
&C_{j,1}, C_{j,2}, C_{j,3}, C_{j,4}),  j = 1, \cdots, L.
\end{align}
\begin{align}
Z_2 = &(A_{j,1}, A_{j,2}, A_{j,5}, A_{j,6}, B_{j,1}, B_{j,2}, B_{j,5}, B_{j,6}, \notag\\
&C_{j,1}, C_{j,2}, C_{j,5}, C_{j,6}), j = 1, \cdots, L.
\end{align}
\begin{align}
Z_3 = &(A_{j,3}, A_{j,4}, A_{j,5}, A_{j,6}, B_{j,3}, B_{j,4}, B_{j,5}, B_{j,6}, \notag\\
&C_{j,3}, C_{j,4}, C_{j,5}, C_{j,6}), j = 1, \cdots, L.
\end{align}
In this example, we consider the demand $\fsf = (A, B, C)$. Since the request vector contains distinct files, 
specifying which segment of each file is requested (i.e., the vector  $\ssf$) is irrelevant and shall be omitted. 
In the coded delivery phase (see Fig.~\ref{abc-user}), user $1$ multicasts $B_{3} + C_{1}$ (useful to both user 2 and 3), 
user $2$ multicasts $A_{5} + C_{2}$ (useful to both users 1 and 3) and user $3$ multicasts 
$A_{6}+B_{4}$ (useful to both users 1 and 2). 
It follows that $R(2) = R_1+R_2+R_3 = \frac{1}{6} \cdot 3 = \frac{1}{2}$ is achievable. 

\begin{figure}
\centering
\includegraphics[width=12cm]{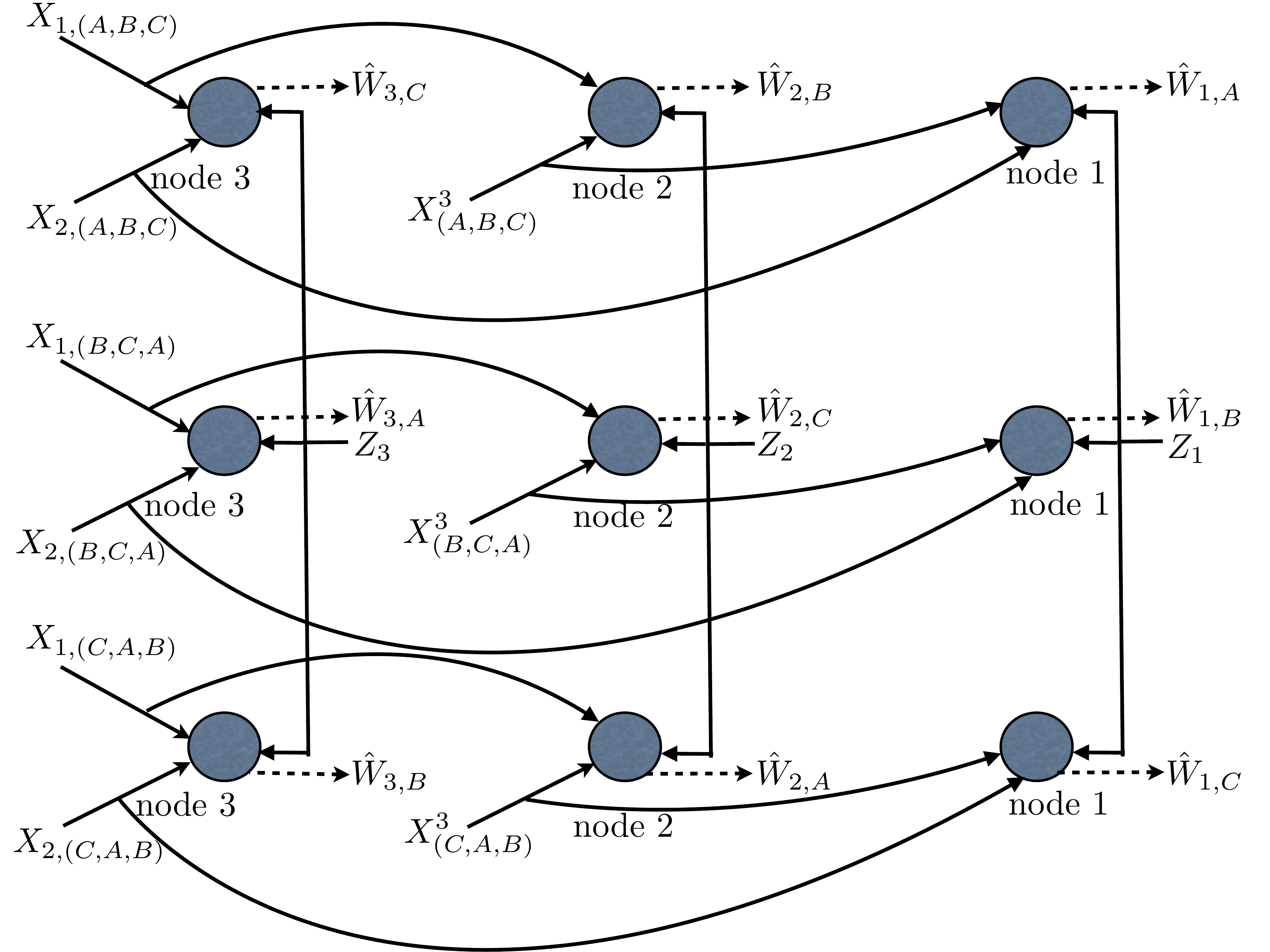} 
\caption{The augmented network when $m=3$, $n=3$. The three requested vectors are: $(A,B,C)$, $(B,C,A)$ and $(C,A,B)$.}
\label{fig: Network_Coding_Model}
\end{figure}

Next, we illustrate the idea of the general rate lower bound of Theorem \ref{theorem: 2}.
Without loss of generality, we assume that $L/L'$ is an integer and let $s$ denote the segment index.
For any scheme that satisfies arbitrary demands $\fsf$, with arbitrary segments $\ssf$, we denote by 
$R_{u,s,\fsf}^{\rm T}$ the number of {\em coded bits}  transmitted  by user $u$, relative to segment $s$ and request 
vector $\fsf$.  
Since the requests are arbitrary, we can consider a compound extension for all possible request vectors. 
For example, we let the first request be $\fsf=(A,B,C)$, the second request be $\fsf = (B, C, A)$ and the third request be $\fsf = (C,A,B)$.
Then, the augmented compound-extended graph is shown in  Fig.~\ref{fig: Network_Coding_Model} where, 
consistently with our general notation defined in Section \ref{sec: Network Model and Problem Formulation},
$Z_u$ denotes the cached symbols at user $u = 1,2,3$, $X_{u,\fsf}$ denotes the transmitted message from user $u$ in correspondence of demand $\fsf$, 
and $\hat{W}_{u,f}$ is the decoded message at user $u$ relative to file $f$. 
Considering user 3, from the cut that separates $(X_{1,(A,B,C)}, X_{2,(A,B,C)}, X_{1,(B,C,A)}, X_{2,(B,C,A)}, X_{1,(C,A,B)}, X_{2,(C,A,B)}, Z_3)$ 
and $(\hat{W}_{3,C}, \hat{W}_{3,A}, \hat{W}_{3,B})$, 
and by using the fact that the sum of the entropies of the received messages and the entropy of the side information (cache symbols) 
cannot be smaller than the number of requested {\em information bits}, we obtain that
\begin{align} \label{eq: converse 1}
&\sum_{s=1}^{\frac{L}{L'}} \left(R_{1,s,(A,B,C)}^{\rm T}+R_{2,s,(A,B,C)}^{\rm T} + R_{1,s,(B,C,A)}^{\rm T} +R_{2,s,(B,C,A)}^{\rm T} \right. \notag\\
&\left.  + R_{1,s,(C,A,B)}^{\rm T}+R_{2,s,(C,A,B)}^{\rm T} \right) + MFL \geq 3FL' \cdot L/L'.
\end{align}
Similarly, from the cut that separates $(X_{1,(A,B,C)}$, $X_{3,(A,B,C)}$, $X_{1,(B,C,A)}$, $X_{3,(B,C,A)}$, $X_{1,(C,A,B)}$,
$X_{3,(C,A,B)}$, $Z_2)$ and $(\hat{W}_{2,B}, \hat{W}_{2,C}, \hat{W}_{2,A})$, 
and from the cut that separates $(X_{2,(A,B,C)}$, $X_{3,(A,B,C)}$, $X_{2,(B,C,A)}$, $X_{3,(B,C,A)}$, $X_{2,(C,A,B)}$,
$X_{3,(C,A,B)}$, $Z_1)$ and $(\hat{W}_{1,A}, \hat{W}_{1,B}, \hat{W}_{1,C})$, 
we obtain analogous inequalities up to index permutations. 
By summing (\ref{eq: converse 1}) and the other two corresponding inequalities and dividing all terms by 2, we obtain 
\begin{align}
&\sum_{s=1}^{\frac{L}{L'}}  \left(R_{1,s,(A,B,C)}^{\rm T} + R_{2,s,(A,B,C)}^{\rm T} + R_{3,s,(A,B,C)}^{\rm T} \right . \notag \\
& + R_{1,s,(B,C,A)}^{\rm T} + R_{2,s,(B,C,A)}^{\rm T}  + R_{3,s,(B,C,A)}^{\rm T}   \notag\\
&\left. + R_{1,s,(C,A,B)}^{\rm T} + R_{2,s,(C,A,B)}^{\rm T} + R_{3,s,(C,A,B)}^{\rm T} \right) + \frac{3}{2}MFL \geq \frac{9}{2} FL. \label{chugabun}
\end{align}
Since we are interested in minimizing the worst-case rate, the sum $R^T_{1,s,\fsf} + R^T_{2,s,\fsf} + R^T_{3,s,\fsf}$ must yields the 
same min-max value $R^{\rm T}$ for any $\ssf$ and $\fsf$. This yields the bound
\be
\frac{3 L}{L'} R^{\rm T} \geq \frac{9}{2} FL  - \frac{3}{2}MFL.
\ee
Finally, by definition of rate $R(M)$, we have that $R(M) = R^T / (FL')$. Therefore, 
dividing both sides of (\ref{chugabun}) by $3FL$, we obtain that the best possible achievable rate must satisfy
\be
R^*(M)  \geq \frac{3}{2} - \frac{1}{2}M.
\ee
In the example of this section, for $M = 2$ we obtain $R^*(2) \geq \frac{1}{2}$. 
Therefore, in this case the achievability scheme given before is information theoretically optimal. 

In the same case of $n=3$ users, per-node storage capacity $M=2$ and library size $m = 3$, 
the coded multicasting scheme of \cite{maddah2012fundamental} where a single codeword is sent to all users through a common bottleneck link 
achieves $R(2) = \frac{1}{3}$.  Then, in this case, the relative loss incurred by not having a base station with access to 
all files is $3/2$.

\subsection{Discussions}

The achievable rate of  \emph{Theorem}~\ref{theorem: 1} can be written as the product of three terms, 
$R(M) = n\left(1-\frac{M}{m}\right)\frac{m}{Mn}$ with the following interpretation: 
$n$ is the number of transmissions by using a conventional scheme that serves individual demands without exploiting the
asynchronous content reuse;  $\left(1-\frac{M}{m}\right)$ can be viewed as the {\em local caching gain}, since any user can cache 
a fraction $M/m$ of any file, therefore it needs to receive only the remaining part;  $\frac{m}{Mn}$ is the {\em global caching gain}, due to the ability
of the scheme to turn the individual demands into a coded multicast message, such that transmissions are useful to many users 
despite the streaming sessions are strongly asynchronous. An analogous interpretation can be given for the terms appearing in the rate
expression achievable with the scheme presented in \cite{maddah2012fundamental} (see Section \ref{section: intro}), 
where the base station has access to all the files. Comparing this rate with our Theorem~\ref{theorem: 1}, we notice that they differ 
only in the last term (global caching gain), which in the base station case is given by $(1+\frac{nM}{m})^{-1}$. For $nM \gg m$, we notice that these 
factors are essentially identical. 

As already noticed, \emph{Theorem}~\ref{theorem: 3} shows that there is no fundamental cumulative gain by using both 
spatial reuse and coded multicasting. Under our assumptions, spatial reuse may or may not be convenient, 
depending whether $\frac{C_r}{\Kc}$ is larger or smaller than $C_{\sqrt{2}}$. A closer look reveals a more subtle tradeoff. 
Without any spatial reuse, let $t=\frac{Mn}{m} \in \mathbb{Z}^+$, the length of the codewords in coded subpackets 
for each user, related to the size of the subpacketization,  is $t{n \choose t}$. This may be very large when $n$ and $M$ are large. 
At the other extreme, we have the case where the cluster size is 
the minimum able to cache the whole library in each cluster. In this case, we can just store $M$ different whole files into each node, 
such that all $m$ files are present in each cluster, and for the delivery phase we just serve whole files without any coding 
as in \cite{ji2013optimal}.  In this case, the achieved throughput is $\frac{C_r}{\Kc} \frac{M}{m}$ bits/s/Hz, 
which is almost as good as the coded scheme, which achieves $\frac{C_r}{\Kc\left(\frac{m}{M}-1\right)}$. 
This simple scheme is a  special case of the general setting treated in this paper, where 
spatial reuse is maximized and codewords have length 1. If we wish to use the achievable scheme of this paper, 
the codewords length is $\frac{M g_c}{m}{g_c \choose \frac{M g_c}{m}}$.
Hence, spatial reuse yields a reduction in the codeword length of the corresponding coded 
multicasting scheme. 

Further, we also notice that even though our scheme is design for a D2D system, it can also be used in a peer-to-peer (P2P) wired network, 
where each peer is allowed to cache information with limited storage capacity. 
By using our approach in the case of $r \geq \sqrt{2}$, peers can exchange multicast messages 
which are useful for a large number of other peers, provided that the network supports multicasting.

\section{Decentralized Random Caching}
\label{sec: A Decentralized Random Caching Construction}

The main drawback of the deterministic caching placement in the achievability strategy of 
Theorem \ref{theorem: 1} is that, in practice, a tight control on the users caches must be implemented in order to
make sure that, at any point in time, the files subpackets are stored into the caches in the required way. 
While this is conceptually possible under our time-scale decomposition assumption 
(see comment in Section \ref{section: remarks}), such approach is not robust to events such as user mobility 
and nodes turning on and off, as it may happen in a D2D wireless network with caching in the user devices.  
In this section, we present a decentralized random caching and coded delivery 
scheme that allows for more system robustness.

\subsection{Transmission range $r \geq \sqrt{2}$}

Decentralized random caching with coded multicast delivery has been considered in \cite{maddah2013decentralized}.
However, there is an important difference between our network model and that of \cite{maddah2013decentralized}, where, 
thanks to the central omniscient server (base station), possibly missing packets due to the decentralized random caching can always be supplied by the 
server by unicasting (or naive multicasting). In our system this is not possible, 
since no node has generally access to the whole file library.  
Hence, in order to ensure a vanishing probability of error, we shall use an additional layer of 
Maximum Distance Separable (MDS) coding and consider the limit of large packet size $F$.

We distinguish between two regimes: $t \eqdef \frac{Mn}{m} > 1$ and $t=1$.  
As already noticed before, $t<1$ is not valid since in this case the file library cannot 
be cached into the network and therefore the worst case user demands cannot be  satisfied.

For $t>1$, we consider the following scheme: 
each file segment of $F$ bits is divided into $K$ blocks of $F/K$ bits each, hereafter referred to as ``subpackets''. 
These subpackets are interpreted as  the elements of the binary extension field $\FF_{2^{F/K}}$, and are encoded using a $(K,K/\rho)$-MDS 
code, for some $\rho < 1$ the choice of which will be discussed later.  Notice that this expands the size of each packet from $F$ to $F/\rho$. 
The resulting $K/\rho$ encoded blocks of $F/K$ bits each will be referred to as ``MDS-coded symbols''. 
The definition of worst-case error probability and achievable rates given in Section \ref{sec: Network Model and Problem Formulation} hold verbatim in 
this case. In particular, since the definition of achievable rate considers a sequence of coding schemes for $F \rightarrow \infty$, 
we can choose $K$ as a function 
of $F$ such that both $K$ and $F/K$ grow unbounded as $F$ increases. This ensures that, for any fixed $\rho$, 
$(K, K/\rho)$-MDS codes exist \cite{lin2004error, yeung2008information}. 
The MDS-coded symbols are cached at random by the user nodes according to Algorithm \ref{algorithm: 5}. 

\begin{algorithm}[ht]
\caption{Decentralized random caching placement scheme}
\label{algorithm: 5}
\begin{algorithmic}[1]
\STATE Encode the $K$ subpackets of each packet of each file by using a $(K, K/\rho)$-MDS code over $\FF_{2^{F/K}}$, 
for some MDS coding rate $\rho < 1$.
\STATE The MDS-coded symbols for each packet of each file are indexed by $\{1, 2, \cdots, K/\rho\}$.
\FORALL{$u = 1, \cdots, n$}
\STATE User $u$, independently of the other users, chooses with uniform probability 
an index set $\Ssf_u$ over all possible sets obtained by sampling without replacement $\frac{M K}{m}$ elements from $\{1, 2, \cdots, K/\rho\}$. 
\STATE User $u$ caches the MDS-coded symbols indexed by $\Ssf_u$ for each packet of each file. 
\ENDFOR
\end{algorithmic}
\end{algorithm}

Next, we describe  a delivery scheme that provides to each requesting user 
enough MDS-coded symbols such that it can recover the desired file segments.
%
For $k = 1, \cdots, L'$,  let $s_u + k - 1$ denote the index of the $k$-th requested 
packet of file $f_u$ by user $u$.  Let $\{Z_{f}^{i,j}: j = 1, \cdots, K/\rho\}$ denote the 
block of MDS-coded symbols from packet $i$ of file $f$.  
For $b = n, n-1, \ldots, 2$, and each user subset $\Usf \subseteq \mathcal{U}$ of size $|\Usf|=b$, 
we define $Z_{f_{u}, \Usf \backslash \{u\}}^{s_u+k-1}$ as the symbol sequence obtained 
by concatenating (in any  pre-determined order) the symbols needed by $u \in \Usf$, 
present in {\em all} the caches of users $v \in \Usf \backslash \{u\}$, 
and not present in the cache of  {\em any} other user $v' \notin \Usf \backslash \{u\}$.
We refer to $Z_{f_{u}, \Usf \backslash \{u\}}^{s_u+k-1}$ as the symbols relative ot $u \in \Usf$ and 
{\em exclusively} cached in all nodes $v \in \Usf \backslash \{u\}$. 
Formally, the index set of the symbols forming $Z_{f_{u}, \Usf \backslash \{u\}}^{s_u+k-1}$ is given by 
\[ J_{f_u, \Usf \backslash \{u\}} = \bigcap_{v  \in \Usf \backslash \{u\}}  
\left\{ j \in \{1 ,\cdots, K/\rho\} : Z_{f_u}^{s_u+k-1,j} \in Z_v \backslash \left \{ 
\bigcup_{u' \notin \Usf \backslash \{u\}} Z_{u'} \right\} \right \}, \]
where the index set does not depend on the packet index $s_u + k - 1$ since the same caching rule 
is used for all packets (see Algorithm \ref{algorithm: 5}). 
Then,  $Z_{f_{u}, \Usf \backslash \{u\}}^{s_u+k-1}$ is the sequence of MDS-coded symbols formed 
by concatenating  the symbols $\{ Z_{f_u}^{s_u+k-1,j} : j \in J_{f_u, \Usf \backslash \{u\}}\}$. 
By construction, each user $v \in \Usf$ has one local replica of $Z_{f_{u}, \Usf \backslash \{u\}}^{s_u+k-1}$ (common symbols) 
for each $u \in \Usf \backslash \{v\}$.  We also let $J_{\Usf}^{\rm max} = \max_{u \in \Usf} |J_{f_u, \Usf \backslash \{u\}}|$, such that 
all sequences  $Z_{f_{u}, \Usf \backslash \{u\}}^{s_u+k-1}$ can be zero-padded\footnote{With a slight abuse of notation we indicate
by $Z_{f_{u}, \Usf \backslash \{u\}}^{s_u+k-1}$ also the zero-padded version.}
to the common maximum length   $J_{\Usf}^{\rm max}$. In order to deliver the MDS-coded symbols, 
each user $v \in \Usf \backslash \{u\}$ sends distinct (i.e., non-overlapping) segments of length $\frac{1}{b-1} \cdot J_{\Usf}^{\rm max} \cdot \frac{F}{K}$
of the sequence of XORed MDS-coded symbols $\bigoplus_{u \in \Usf, u \neq v} Z_{f_{u}, \Usf \backslash \{u\}}^{s_u+k-1}$.
The delivery phase is summarized in Algorithm \ref{algorithm: 6}. 

\begin{algorithm}[ht]
\caption{Decentralized random caching delivery scheme}
\label{algorithm: 6}
\begin{algorithmic}[1]
\FORALL{$k = 1 ,\cdots, L'$}
\FORALL {$b = n, n-1, \cdots, 2$}
\FORALL{$\Usf \subset \mathcal{U}$ with $| \Usf | = b$}
\STATE $J_{\Usf}^{\rm max} = \max_{u \in \Usf} |J_{f_u, \Usf \backslash \{u\}}|$
\FORALL{$v \in \Usf $}
\STATE User $v$ transmits a non-overlapping segment of length 
$\frac{1}{b-1} \cdot J_{\Usf}^{\rm max} \cdot \frac{F}{K}$ of the zero-padded and XORed MDS-coded symbol sequence 
$\bigoplus_{u \in \Usf, u \neq v} Z_{f_{u}, \Usf \backslash \{u\}}^{s_u+k-1}$.  
\ENDFOR
\ENDFOR
\ENDFOR
\ENDFOR
\end{algorithmic}
\end{algorithm}

\begin{figure}
\centering
\subfigure[]{
\centering \includegraphics[width=10cm]{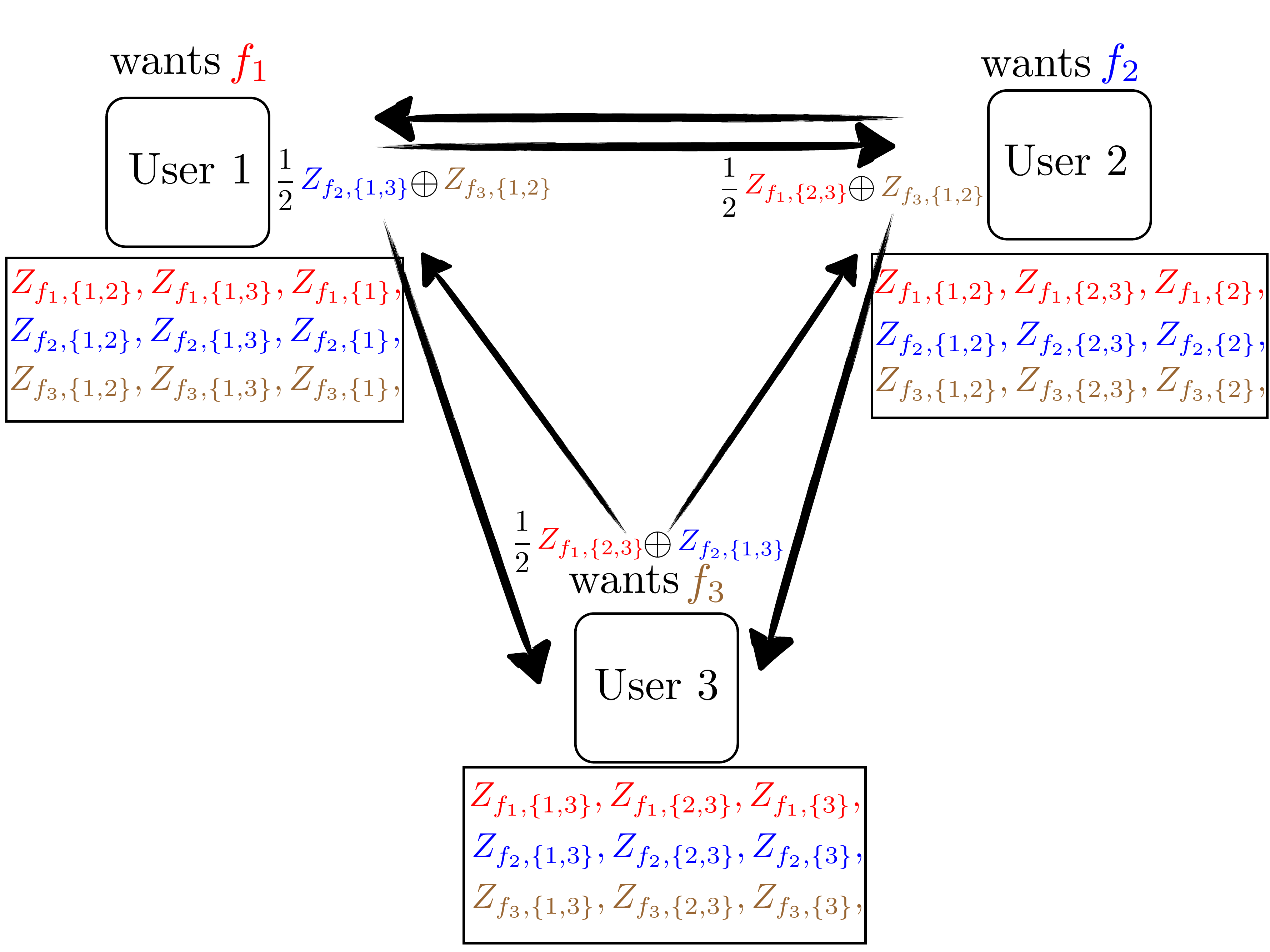}
\label{fig: example_2_1}
}
\subfigure[]{
\centering \includegraphics[width=10cm]{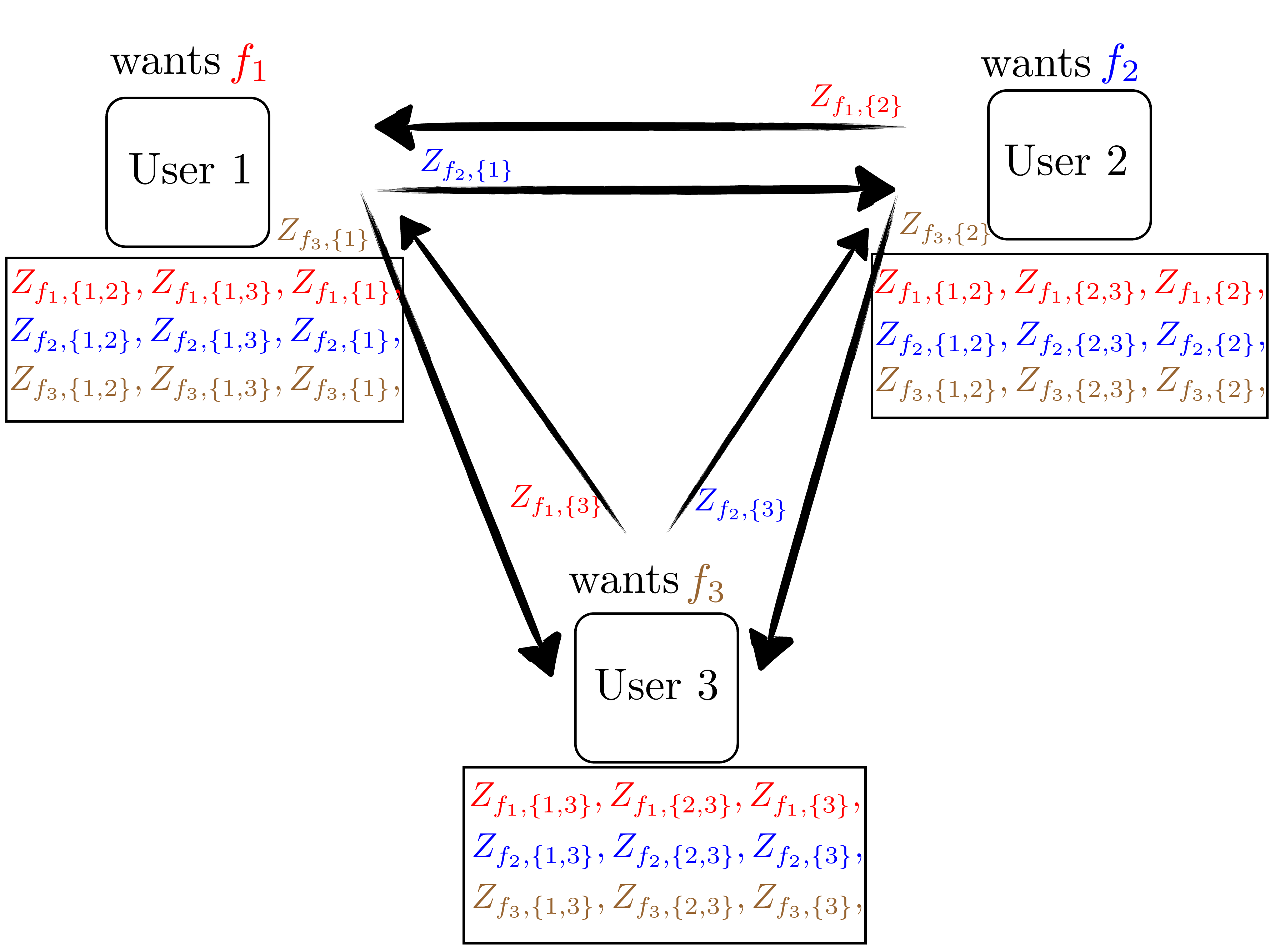}
\label{fig: example_2_2}
}
\caption{Illustration of the example of three users with $M = 2$, $m=3$, achieving rate $R(2) = 0.77$, where $\rho = 0.95$ ($\rho^*=0.9510$, $\varepsilon$ is chosen as $0.001$.). a)~ The first iteration with $|\Usf|=b=3$. b)~ The second iteration with $|\Usf|=b=2$.}
\label{fig: example_two}
\end{figure}

Since the scheme is admittedly complicated and its general description relies on a heavy notation, 
we provide here an illustrative example (see Fig.~\ref{fig: example_two}). 
As in the case of deterministic caching, consider the case of 3 users denoted by 1,2,3.  
Neglecting the packet superscript (irrelevant in this example), {for the first round of the scheme, with $b =3$ 
(see Fig. \ref{fig: example_2_1}),} let $Z_{f_1, \{2,3\}}$ be the sequence of MDS-coded symbols 
useful to user 1 (requesting file $f_1$) and present in the caches of users 2 and 3.
Also, let $Z_{f_2, \{1,3\}}$ and $Z_{f_3, \{1,2\}}$ have similar and corresponding meaning, after permuting the indices. 
Then, user 1 forms the XORed sequence $Z_{f_2, \{1,3\}} \oplus Z_{f_3, \{1,2\}}$, 
user 2 forms the XORed sequence $Z_{f_1, \{2,3\}} \oplus Z_{f_3, \{1,2\}}$, and
user 3 forms the XORed sequence $Z_{f_1, \{2,3\}} \oplus Z_{f_2, \{1,3\}}$. Finally, each user transmits to the other two users
$1/2$ of its own XORed sequence. {For the second round of the scheme, with $b = 2$ (see Fig. \ref{fig: example_2_2}), 
let $Z_{f_1, \{2\}}$ and $Z_{f_1, \{3\}}$ denote the sequence of MDS-coded symbols useful to user 1 and cached exclusively by user 2 and user 3, respectively. Similarly, let $Z_{f_2, \{1\}}$, $Z_{f_2, \{3\}}$, $Z_{f_3, \{1\}}$ and $Z_{f_3, \{2\}}$ have corresponding meaning. 
Since $b = 2$, there is no multicasting opportunity. User 1 will just transmit sequence $Z_{f_2, \{1\}}$ to user 2 and $Z_{f_3, \{1\}}$ to user 3. 
Users 2 and 3 perform similar operations.} 
Focusing on decoding at user 1, {after the first round}, half of $Z_{f_1,\{2,3\}}$ is recovered from user  2 
transmission  and the other  half from user 3 transmission, by using the side information of its own cache. 
{After the second round, $Z_{f_1, \{2\}}$ is directly received from user 2, and 
$Z_{f_1, \{3\}}$ from user 3.  Finally,  if the MDS coding rate $\rho$ is chosen appropriately, 
user 1 is able to recover the desired file $f_1$ with high probability
from the MDS-coded symbols $Z_{f_1,\{2,3\}}, Z_{f_1,\{2\}}$ and $Z_{f_1,\{3\}}$ 
and the symbols relative to file $f_1$ already present in its cache.} 

The following result yields a sufficient condition for the MDS coding rate $\rho$ such that, in the general case, 
all files can be decoded with high probability from the MDS coded symbols  cached in the network:

\begin{theorem}
\label{theorem: caching}
Let $\rho = (1-\varepsilon)\rho^*$, where $\varepsilon > 0$ is an arbitrarily small constant and $\rho^*$ is the non-zero solution of the fixed 
point equation:
\be
\label{eq: W}
x = 1 - \exp(-tx).
\ee
Then, the random caching scheme of Algorithm \ref{algorithm: 5} with MDS coding rate $\rho$ yields, for all $f = 1, \ldots, m$,
\begin{eqnarray}
\label{eq: prob MDS}
&& \PP(\text{File $f$ can be decoded from the cached MDS-coded symbols in the network}) \notag\\
&& \geq 1 - \exp\left(-K^{\delta_1(\varepsilon)} + o\left(K^{\delta_1(\varepsilon)}\right) \right),
\end{eqnarray}
where $\delta_1(\varepsilon)$ is a term independent of $K$, {such that $\delta_1(\varepsilon) > 0$ for all $\varepsilon > 0$.}
\hfill  $\square$
\end{theorem}
Theorem \ref{theorem: caching} is proved in Appendix \ref{sec: Proof of theorem caching}.
From Theorem \ref{theorem: caching} and the union bound, we have immediately that all the files can be 
successfully decoded from the cached MDS-coded symbols in the network with 
arbitrarily high probability for sufficiently large $K$. 

{In our example, i.e., for $n=3$, $m=3$ and $M=2$, by choosing $\varepsilon$ in Theorem \ref{theorem: caching} as $0.001$, 
we obtain $\rho = 0.95$, yielding an achievable rate $R(2) = 0.77$ (see (\ref{eq: decentralized 1}) in Theorem \ref{theorem: decentralized 1}).}  

When $t = 1$, the scheme based on MDS codes given by Algorithms \ref{algorithm: 5} and \ref{algorithm: 6} cannot 
be applied since (\ref{eq: W}) has no finite positive solution. 
In this case, we propose a different caching and delivery scheme given as follows.
Each packet of each file is divided into $K$ subpackets of size $F/K$ bits, 
and each block of $K$ subpackets, interpreted as symbols over $\FF_{2^{F/K}}$, 
is separately and independently encoded by each user $u$ by using a random linear 
{\em hashing function} that compresses the $K$ symbols into $MK/m$ symbols as follows:
each user $u$ generates independently a matrix $\Gm_u$ of dimension $K \times MK/m$  over $\FF_{2^{F/K}}$ 
with i.i.d. components. Representing $W_f^i$ (the $i$-th packet of file $f$)
as a $1 \times K$ vector $\wv_f^i$  over $\FF_{2^{F/K}}$, the hashing transformation at user $u$ is given by $\cv_{f,u}^i = \wv_f^i \Gm_u$.  
Then, each user $u$ caches $\cv_{f,u}^i$ for all $f = 1, \ldots, m$ and $i = 1,\ldots, L$. Notice that, 
in this way, the sum of the lengths of these codewords is $Lm \times MK/m \times F/L = LMF$ bits, such that the cache size constraint is satisfied
with equality.

For the delivery phase, each user unicasts $\frac{1}{n-1}\left(1-\frac{M}{m}\right)K$ coded symbols 
for each requested packet of all other users. Hence, at the end of the delivery phase, each 
requesting user collects $\left(1-\frac{M}{m}\right)K$ coded symbols from the other $n - 1$ users and $MK/m$ symbols from its own 
hashed codeword, such that it has a total of $K$ coded symbols. If the $K \times K$ system of linear equations corresponding 
to these symbols has rank $K$, then the packet can be retrieved. This condition is verified with arbitrarily large probability for sufficiently large $F/K$ 
\cite{ho2006random, yeung2008information}. Furthermore, we observe that the caching and delivery scheme for 
$t = 1$ can be applied, trivially, also for $t >1$.  Eventually, combining the two caching and delivering schemes, 
we can prove the following general results: 

\begin{theorem}
\label{theorem: decentralized 1}
For $r  \geq \sqrt{2}$ and $t  = \frac{Mn}{m} > 1$, as $F,K \rightarrow \infty$ with sufficiently large ratio $F/K$, the following rate is achievable
by decentralized caching:
\begin{equation}
R(M) = \min\left\{\frac{1}{\rho}\sum_{s=2}^n {n \choose s}\frac{s}{s-1}\left(\frac{M\rho}{m}\right)^{s-1}\left(1-\frac{M\rho}{m}\right)^{n-s+1}, n-t\right\}. 
\label{eq: decentralized 1} 
\end{equation}
Consequently, the throughput  $T(M) = \frac{C_{\sqrt{2}}}{R(M)}$ is also achievable. 
\hfill  $\square$
\end{theorem}

In order to evaluate the achievable rate of Theorem \ref{theorem: decentralized 1}, the following lemma is useful:

\begin{lemma} \label{upper bound decentralized}
The achievable rate $R(M)$ of Theorem \ref{theorem: decentralized 1} is upper bounded by 
\begin{align}
R(M)  \leq & \min\left\{ \frac{m}{M\rho^2}\left(1-\frac{M\rho}{m}\right)  \left(1+\frac{3}{(n+1)\frac{M\rho}{m}} \left(1-\left(1-\frac{M\rho}{m}\right)^{n+1}\right) \right . \right . \notag \\
& \left .  \left . - 4\left(1-\frac{M\rho}{m}\right)^n - \frac{5}{2}\frac{M\rho n}{m}\left(1-\frac{M\rho}{m}\right)^{n-1}\right), n-t\right\}. \label{eq: decentralized 2} 
\end{align}
\hfill  $\square$
\end{lemma}

The proof of Theorem \ref{theorem: decentralized 1} and of Lemma \ref{upper bound decentralized} are 
given in Appendix \ref{sec: Proof of Theorem theorem: decentralized 1}. For $t=1$, we have:

\begin{corollary}
\label{theorem: decentralized 2}
For $r \geq \sqrt{2}$ and $t  = \frac{Mn}{m} = 1$, as $F,K \rightarrow \infty$ with sufficiently large ratio $F/K$, the following rate is achievable:
\be
R(M) = \frac{m}{M}\left(1 - \frac{M}{m}\right).
\ee
\hfill  $\square$
\end{corollary}
Corollary \ref{theorem: decentralized 2} is immediately obtained by using the second term in the ``min'' of (\ref{eq: decentralized 1}) 
and letting $t=1$.  

The gap between the achievable rate and the lower bound of Theorem \ref{theorem: 2}, which applies to any scheme, also centralized, 
is given by: 

\begin{theorem}
\label{theorem: decentralized gap}
For $r  \geq \sqrt{2}$ and $t  = \frac{Mn}{m} \geq 1$, as $F,K \rightarrow \infty$ with sufficiently large ratio $F/K$, then let $m,n \rightarrow \infty$, for any $M \leq \frac{1}{1 + \varepsilon}m$, for some arbitrarily small $\varepsilon>0$, 
the ratio of the achievable rate with decentralized caching over the optimal rate 
with unrestricted caching is upper bounded by
\be
\label{eq: decentralized gap 3}
\frac{R(M)}{R^*(M)} \leq \left\{\begin{array}{cc}\frac{8}{(1-\varepsilon)^2}, &   t = \omega(1), \frac{1}{2} \leq M = o(m) \\ 
6, & M = \Theta(m)\\
\frac{4}{M(1-\varepsilon)^2}, & n = \omega(m), M <\frac{1}{2} \\
\min\{4t, f_g(t)\}, & \text{Otherwise} 
\end{array}\right.,
\ee
where 
\be
f_g(t) = \frac{1}{\rho^2}  \left(1 + f_ \rho(t)\right) \left\{\begin{array}{cc}\frac{4t}{\lfloor t \rfloor}, &  n = O(m), t=\Theta(1), \frac{1}{2} \leq M = o(m) \\
\frac{t}{\lfloor t \rfloor}\frac{2}{M}, & n = O(m), n > m, M <\frac{1}{2} \\
2, & n = O(m), n \leq m, M <\frac{1}{2}
\end{array}\right.,
\ee
where $\frac{t}{\lfloor t \rfloor} \leq 2$ and 
$f_\rho(t) =  \frac{3}{\rho t} - e^{-\rho t}\left(\frac{3}{\rho t} + 4 + \frac{5}{2}\rho t\right)$. Further, for $t \neq 1$, $\frac{1}{\rho^2}  \left(1 + f_ \rho(t)\right) $ 
can be upper bounded by a positive constant.
\hfill  $\square$
\end{theorem}
The proof of Theorem~\ref{theorem: decentralized gap} is given in Appendix \ref{sec: proof theorem decentralized gap}. The gap between $T(M)$ and $T^*(M)$ follows as a consequence.

When naive multicasting is allowed, if $t>1$, with high probability (given by (\ref{eq: prob MDS}) in Theorem \ref{theorem: caching}), 
there are at least $K$ distinct coded symbols for each packet of each file cached it the network. Therefore, by requiring multicasting of at most $K$ 
distinct coded symbols for each file, a rate $m$ can be achieved. 
Similarly, when $t=1$, by using the linear random hashing scheme,  there are again at least $K$ distinct coded symbols for each packet of each file 
cached in the network with high probability as the field size ($2^{F/K}$) grows large. 
Then, we can achieve a rate of $m$ by naive multicasting at most $K$ distinct coded symbols 
of each requested packet such that all the users can decode.  Hence, similar to Theorem \ref{theorem: decentralized gap}, 
(\ref{eq: decentralized gap 3}) becomes a constant when naive multicasting is allowed.

\subsection{Transmission range $r < \sqrt{2}$}

Based on the scheme developed for the case $r \geq \sqrt{2}$, by using the same clustering approach for the deterministic caching case, 
we immediately  have:
\begin{corollary}
\label{theorem: decentralized 3}
Let $r$ is such that  any two nodes in a ``squarelet'' cluster of size $g_c$  can communicate, 
as $F,K \rightarrow \infty$ with sufficiently large ratio $F/K$, the throughput
\be
T(M) = \frac{C_r}{\Kc} \frac{1}{R(M)},
\ee
is achievable with decentralized caching,   $\Kc$ is the clustering scheme 
reuse factor, and $R(M)$ is given by \emph{Theorem}~\ref{theorem: decentralized 1} for $t >1$ and
by \emph{Theorem}~\ref{theorem: decentralized 2} for $t =1$.
\hfill  $\square$
\end{corollary}
Furthermore, similar to Theorem \ref{theorem: gap 2}, the ratio $T^*(M)/T(M)$ is upper bounded by  
the terms in 
(\ref{eq: decentralized gap 3}), multiplied by the geometry 
factor  $\Kc \left\lceil \frac{4(2+\Delta)^2}{\Delta^2}\right \rceil$.

\subsection{Discussions}

From the above results, we observe that the proposed decentralized random caching scheme achieves a performance very close to 
that of the deterministic caching scheme.  Specifically, in the case of $r \geq \sqrt{2}$ and $t > 1$, 
from Theorem~\ref{theorem: decentralized gap}, 
we see that our decentralized approach achieves 
order optimality (constant multiplicative gap)  in the scaling throughput law for  large networks, i.e., in the limit of $n \rightarrow \infty$. 
It is important to notice the order in which we have to take the limits here:
for any finite $n,m,M$, we consider the limit for large file size (in packets) $L \rightarrow \infty$, and 
large bits per packet $F \rightarrow \infty$. Then, we look at the rate behavior for possibly large network size $n$ and library size $m$. 
Taking limits in this order is meaningful if we consider typical applications of caching for video on-demand delivery. 
Consider for example a good-quality movie file encoded at 2 MB/s, of total duration of 1h. 
In current Dynamic Adaptive Streaming over HTTP (DASH) standards  \cite{sanchez2011idash, keller2012microcast} 
a video packet has typical duration of 1s, corresponding to
2 Mb. In this case we would have $L = 3600$ and $F = 2 \cdot 10^6$, which justify our assumptions. 

\begin{figure}
\centering
\subfigure[]{
\centering \includegraphics[width=7.5cm]{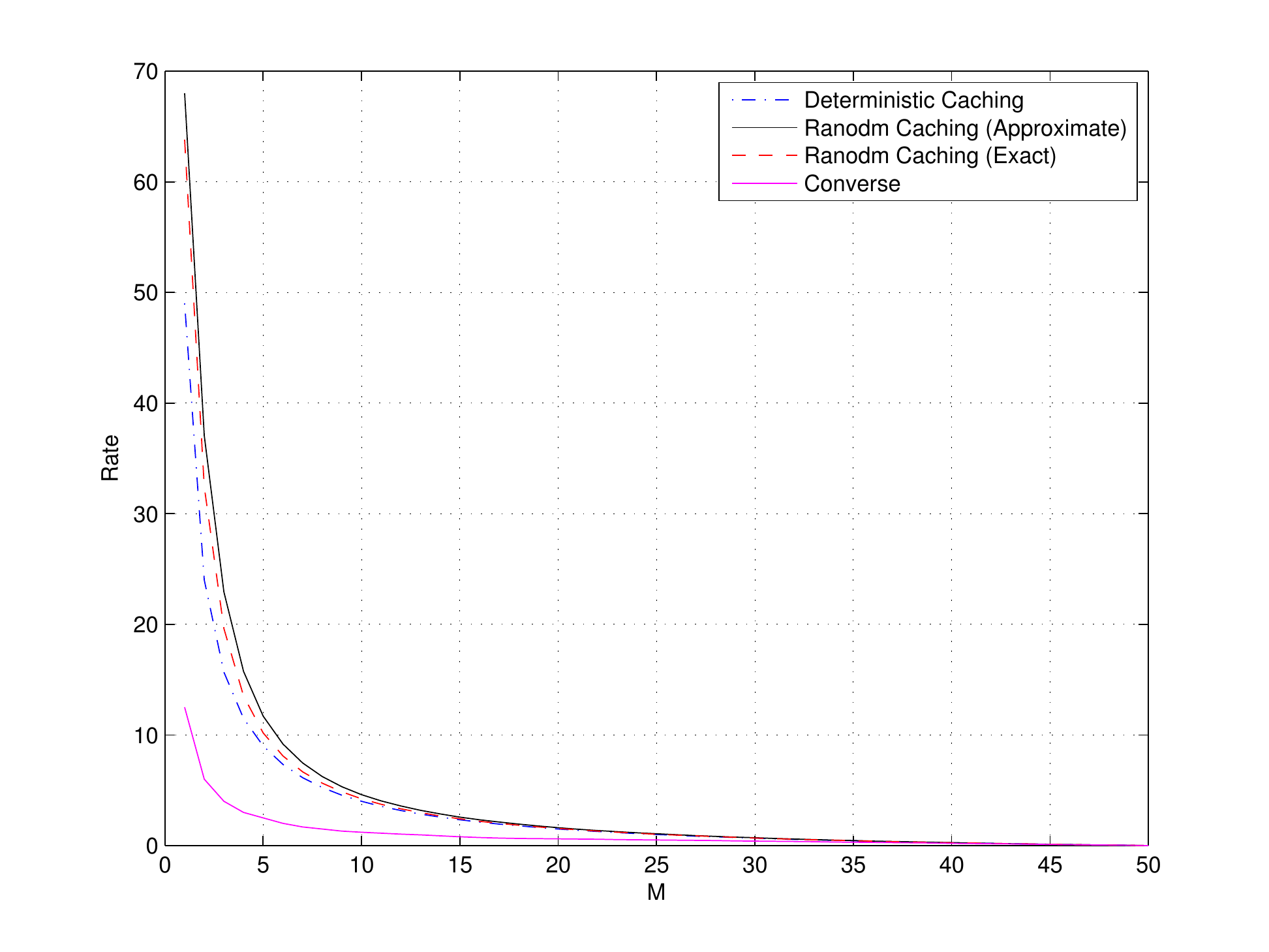}
\label{fig: result_cen_decen_2}
}
\subfigure[]{
\centering \includegraphics[width=7.5cm]{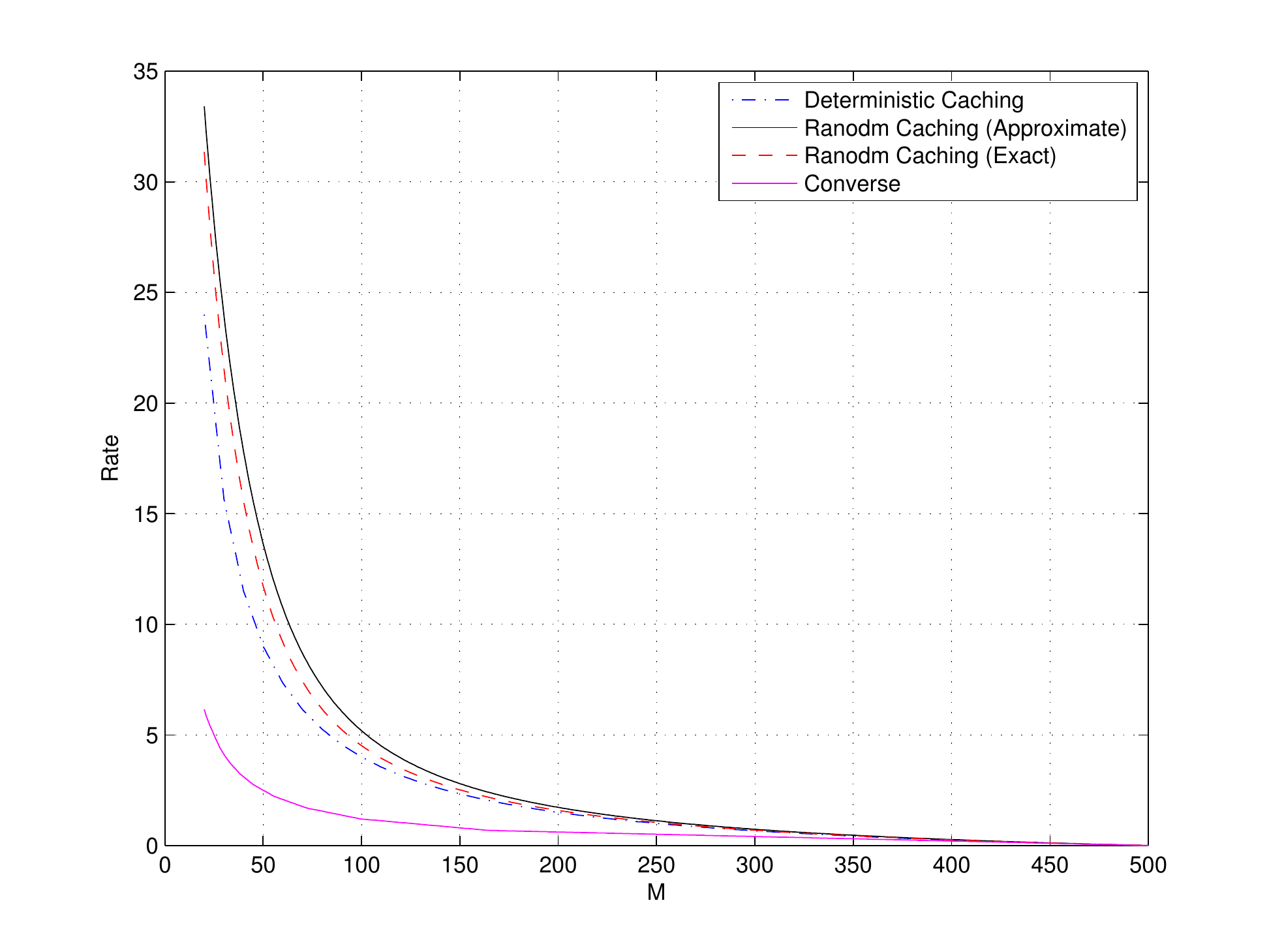}
\label{fig: result_cen_decen_1}
}
\caption{Rate $R(M)$ as a function of the cache size $M$ for deterministic and random D2D caching.
In (a) we let $m=50$ and $n=100$. In (b) we let $m=500$ and $n=50$.
The rate of deterministic caching 
is given by (\ref{eq: theorem 1}). The rate of 
Random Caching (Exact) is given by (\ref{eq: decentralized 1}). 
The curve ``Random Caching (Approximate)'' is plotted by using (\ref{eq: decentralized 2}). The converse of the rate is given by (\ref{banana}). 
}
\end{figure}

For finite $n$, from Theorem \ref{theorem: decentralized gap} and its proof, we can see that the multiplicative gap between the achievable rate 
of the decentralized random caching approach and that of the centralized deterministic caching approach 
is a function of the system parameters $M$,$m$ and $n$. However, from the simulation 
results (see Figs.~\ref{fig: result_cen_decen_2} and \ref{fig: result_cen_decen_1}), 
we observe that this gap vanishes as the memory size $M$ increases. 
In addition, from Theorem \ref{theorem: decentralized gap}, as $n,m \rightarrow \infty$,  this gap becomes a constant. 
Hence, the  decentralized random caching scheme performs approximately as well as the centralized 
deterministic caching scheme in the most interested regimes of the system parameters. 


\section{Conclusions}
\label{sec: Conclusions}

In this paper, we have determined constructive achievability coding strategies and information theoretic bounds 
for a D2D caching network under the constraint of arbitrary (i.e., worst-case) demands. 
We have considered two caching and (inter-session network coded) delivery schemes: the first is based on deterministic (centralized) caching, 
and the second is based on random (decentralized) caching. The decentralized nature of the second scheme lies in the fact that
each user independently determines the (coded) symbols to cache, without knowing what the others do. 
Our work differs from concurrent and previous recent works by the fact that we do not consider 
a central omniscient server that has access to all files in the library. Hence, 
the proposed schemes are strictly peer-to-peer and infrastructureless, and therefore
they are suited to a wireless D2D network \cite{wu2010flashlinq , camps2012device, doppler2009device, ji2013wireless}. 

In the case where all nodes in the network are in the reach of each other (range $r \geq \sqrt{2}$, under our normalizations), 
under the assumption of asynchronous content reuse, i.e., when naive multicasting is forbidden or useless by our model, 
we showed that the deterministic caching scheme is optimal within a constant multiplicative factor 
in almost all system regimes, with the exception of the regime of large content reuse (number of users $n$ larger than the number of library files  $m$) 
and very small cache capacity $M < 1/2$. This regime is arguably not very interesting for applications, since the goal of caching is precisely to
trade cache memory in the user devices (an inexpensive and largely untapped network resource) for bandwidth (a very expensive and scarce commodity). 
In any case, allowing for naive multicasting fills also this small gap. 

Interestingly and somehow counterintuitively, we found that when we restrict the transmission range to some $r < \sqrt{2}$ in order to allow  
multiple concurrent transmissions in the network under the protocol model (spatial spectrum reuse), 
the throughput does not improve in terms of the scaling law with respect to  $n,m$ and $M$. 
Spatial reuse has therefore only a possible gain in terms of actual rates, due to the possible improvement of the actual transmission rates due to
the shorter link distance. It follows that, in order to assess whether spatial reuse is beneficial or not,  
one has to consider an accurate model for the underlying physical layer and propagation channel, and consider actual transmission rates and
interference. Evidently, the protocol model considered in this work is too ``coarse'' to capture these aspects. 
An example of such analysis is provided in  \cite{ji2013wireless} for a D2D network with realistic propagation channel modeling 
operating at various frequency bands, from  cellular microwave to mm-wave bands, as envisaged in the 
forthcoming 5G standardization \cite{rappaport2011broadband, azar201328, daniels201060, akdeniz2013millimeter, rangan2013energy, rangan2014millimeter}. 
Moreover, for the deterministic caching scheme, the trade-off between coded multicasting 
and spatial reuse is reflected by the code length, which depends on the communication cluster size. 

In the proposed decentralized random caching scheme, we have used MDS coding (or random linear intra-session network coding in the form of 
random linear hashing of the files subpackets)  in order to ensure, with high probability,  that all files can be recovered 
by the (coded) symbols cached into the network. 
This was not necessary in the setting of \cite{maddah2013decentralized}, since the missing symbols can be always supplied by the omniscient 
broadcasting node.  We showed that also this random caching scheme achieves 
order-optimality when the network size $n$ become large. Overall, the decentralized random caching scheme appears to be more attractive
for practical applications since it allows all users to cache at random and independently their assigned fraction of coded
symbols of the library files, without knowing a priori which symbols have been already cached  by other nodes.  

As a final remark, we wish to stress the fact that a decentralized D2D caching scheme may effectively provide a very attractive
avenue for efficient content distribution over wireless networks, avoiding the cluttering of 
the cellular infrastructure. For example, consider the results of Figs.~\ref{fig: result_cen_decen_2} and \ref{fig: result_cen_decen_1}.
In both cases, when $M \approx m/10$ (i.e., 10\% of the whole library is cached in each user device), 
the $n$ user requests can be satisfied by sending the coded equivalent of 10 files. 
This means that if the physical layer link peak rate is (say) a modest 20 Mb/s, 
each user can stream video at 2 Mb/s,  irrespectively of the number of users. 

\appendices

\section{Deterministic Caching and Delivery Schemes: General Case,  $r \geq \sqrt{2}$}
\label{Caching Placement and Delivery Scheme}

In this section, we generalize the the deterministic caching and coded delivery 
scheme illustrated in Section \ref{sec:example} through an example to the general case of any $m$, $n$ and $M$, such that 
$t \eqdef \frac{Mn}{m} \geq 1$ is an positive integer. When $t$ is not an integer, we can use a resource sharing scheme as in the examples
at the end of this section (see also \cite{maddah2012fundamental,maddah2013decentralized}).

\begin{itemize}
\item {\bf Cache Placement:}
The cache placement scheme is closely related to the scheme in \cite{maddah2012fundamental}. 
Recall that $\mathcal{U} = \{1, 2, \cdots, n\}$ denotes the set of user nodes and 
$W_f^i$ denotes packet $i$ of file $f$.  We divide each packet of each file into $t{n \choose t}$ subpackets. 
Letting $\Tsf$ denote a specific combination of $t$ out of $n$ elements, we index each subpacket by the pair $(\Tsf, j)$ with
$j = 1,\ldots, t$, such that the subpackets of $W_f^i$ are indicated by $\{ W_f^{i,\Tsf,j}\}$. 
Node $u$ caches all the subpackets such that $u \in \Tsf $, for all $f = 1,\ldots, m$ and $i = 1,\ldots, L$, such that 
the cache function $Z_u$ is just given by the concatenation of this collection of subpackets. 

\item {\bf Delivery and Decoding:}
For the delivery phase, let $k = 1, \cdots, L'$ and denote $s_u + k - 1$ as the index of the $k$-th packet of file $f_u$ requested by user $u$. 
As a consequence of the caching scheme described above, 
any nodes subset of size $t+1$ in $\Uc$ has the property that the nodes of any of its subsets of 
size $t$ share $t$ subpackets for every packet of every file. Consider one of these subsets, and consider the remaining $(t+1)$-th node. 
For any file requested by this node, by construction, there are $t$ subpackets shared by all other $t$ nodes 
and needed by  the $(t+1)$-th node. Therefore, each node in every subset of size $t+1$ 
has $t$ subpackets, each of which is useful for one of the remaining $t$ nodes. 
Furthermore, such sets of subpackets are disjoint (empty pairwise intersections). 
For delivery, for all subsets of $t+1$ nodes, each node computes the XOR of its set  of $t$ useful 
subpackets and multicasts it to all other nodes. 
In this way, for every multicast transmission exactly $t$ nodes will be able to decode a useful packet using 
``interference cancellation'' based on their cache side information. 

\end{itemize}

In order to illustrate the caching and delivery scheme described above, we consider a few examples.

\begin{example}
\label{example: 1}
Consider a network with $n=2$, $m=2$ and $M=1$, with $t = 1$. The two files are denoted by $W_1$ and $W_2$. 
First, we divide each packet into $t {n \choose t} = 2$ subpackets.  In this case,  
$\Tsf \in \{\{1\}, \{2\}\}$. Hence, the subpacket labeling is $W_1 = \{(W_1^{i,\{1\},1}, W_1^{i,\{2\},1}) : i = 1,\ldots, L\}$
and $W_2 = \{(W_2^{i,\{1\},1}, W_2^{i,\{2\},1}) : i = 1,\ldots, L\}$. The caches are given by:
\begin{eqnarray*}
Z_1 & = & \{ W_1^{i,\{1\},1},  W_2^{i,\{1\},1} : i = 1,\ldots, L\} \\
Z_2 & = & \{ W_1^{i,\{2\},1},  W_2^{i,\{2\},1} : i = 1,\ldots, L\} 
\end{eqnarray*}
Assuming that, without loss of generality, user $1$ requests  packets $[s_1:s_1+L'-1]$ of file 
$W_1$ and users $2$ requests packets $[s_2:s_2+L'-1]$ of file  $W_2$, 
User $1$ sends $W_2^{i,\{1\},1} : i = s_2, \ldots, s_2+L'-1$ to user 2, 
and user $2$ sends $W_1^{i,\{2\},1} : i = s_1, \ldots, s_1+L'-1$ to user 1.  The transmission rate is 
$R(1) = 2 \times \frac{1}{2} = 1$ (recall that the rate is expressed in number of equivalent transmissions of blocks of $F$ bits). 
\hfill $\lozenge$
\end{example}

\begin{example}
\label{example: 2}
Consider the example of Section \ref{sec:example} expressed in the general notation. We have $n = m = 3$ and $M = 2$, yielding $t = 2$.
Packets are divided into $t{n \choose t} = 6$ subpackets, with the following labeling: for $f = 1,2,3$, let
$W_f = \{(W_f^{i,\{1,2\},1}, W_f^{i,\{1,2\},2},W_f^{i,\{1,3\},1}, W_f^{i,\{1,3\},2},W_f^{i,\{2,3\},1}, W_f^{i,\{2,3\},2}) : i = 1,\ldots, L\}$. 
The caches are given by:
\begin{eqnarray*}
Z_1 & = & \{ W_f^{i,\{1,2\},1},  W_f^{i,\{1,2\},2},W_f^{i,\{1,3\},1},  W_f^{i,\{1,3\},2} : i = 1,\ldots, L, f = 1,2,3 \} \\
Z_2 & = & \{ W_f^{i,\{1,2\},1},  W_f^{i,\{1,2\},2},W_f^{i,\{2,3\},1},  W_f^{i,\{2,3\},2} : i = 1,\ldots, L, f = 1,2,3 \} \\
Z_3 & = & \{ W_f^{i,\{1,3\},1},  W_f^{i,\{1,3\},2},W_f^{i,\{2,3\},1},  W_f^{i,\{2,3\},2} : i = 1,\ldots, L, f = 1,2,3 \}. 
\end{eqnarray*}
Assuming, without loss of generality, that user $u$ requests packets $[s_u:s_u+L'-1]$ from file $W_u$, for $u = 1,2,3$ and some arbitrary segment indices 
$s_1,s_2,s_3$, we apply now the delivery scheme according to the general recipe described above. 
We have a single subset of size $t + 1 = 3$, namely $\{1,2,3\}$. Each user $u$ has $t$ subpackets useful for the other two users,
and such that the sets of such subpackets are disjoint. The choice of the sets is not unique. For example, the following choice of coded multicast messages 
is possible:
\begin{eqnarray*}
X_{1,\{1,2,3\}} & = & W_2^{s_2+i,\{1,3\},1} \oplus W_3^{s_3+i,\{1,2\},1}, \;\; i = 1,\ldots, L'-1 \\
X_{2,\{1,2,3\}} & = & W_1^{s_1+i,\{2,3\},1} \oplus W_3^{s_3+i,\{1,2\},2}, \;\; i = 1,\ldots, L'-1 \\
X_{3,\{1,2,3\}} & = & W_1^{s_1+i,\{2,3\},2} \oplus W_2^{s_2+i,\{1,3\},2}, \;\; i = 1,\ldots, L'-1.
\end{eqnarray*}
As already given in Section \ref{sec:example}, the rate in this case is $R(M) = 3 \times \frac{1}{6} = \frac{1}{2}$. 
\hfill $\lozenge$
\end{example}

\begin{example} \label{example: 3}
This example illustrates the strategy when $t$ is not an integer. In this case, we use a cache sharing scheme 
achieving the lower convex envelope of the rates corresponding to the two integer values $\lfloor t \rfloor$ and $\lceil t \rceil$. 
Consider a network with $n=2$, $m=3$ and $M=2$, yielding $t = Mn/m = 4/3$, between $1$ and $2$. 
For $t = 1$, $m = 3$ and $n = 2$, we obtain $M_1 = 3/2$. For 
$t = 2$, $m = 3$ and $n = 2$, we obtain $M_2 = 3$. Hence, the cache sharing scheme uses a 
fraction $\alpha$ such that $\alpha M_1 + (1 - \alpha)M_2 = M = 2$, yielding $\alpha = 2/3$. 
We allocate $3/2 \cdot 2/3 = 1$ storage capacity to the caching placement for $M_1 = 3/2$ and  $3 \cdot 1/3 = 1$ 
storage capacity to the caching placement for $M_2 = 3$. The details are in the following.

We divide each packet of each library file $W_f$, $f = 1,2,3$ into $2$ subpackets with size $\alpha F$ and$(1 - \alpha) F$, respectively. 
Since $\alpha = 2/3$, we denote the resulting packets of $W^i_f$  as  $W^i_{\{f,\frac{2}{3}\}}$ and $W^i_{\{f,\frac{1}{3}\}}$.
Then, the packets $W^i_{\{f,\frac{2}{3}\}}$ are stored according to the scheme for $M_1 = 3/2$, $t = 1$. 
In particular, each $W^i_{\{f,\frac{2}{3}\}}$ is divided into $t{n \choose t} = 2$ subpackets with $\Tsf \in \{\{1\}, \{2\}\}$. 
For $f = 1, 2, 3$, the subpacket labeling is 
\be
W^i_{\{f,\frac{2}{3}\}} = \left(W_{\{f,\frac{2}{3}\}}^{i,\{1\},1}, W_{\{f,\frac{2}{3}\}}^{i,\{2\},1} \right ) \;\;\; : \;\;\; i = 1, \cdots, L.
\ee
Similarly, the packets $W^i_{\{f,\frac{1}{3}\}}$ are stored according to the scheme for $M_2 = 3$, $t=2$, with
$t {n\choose t} = 2$ subpackets and $\Tsf = \{1,2\}$.  For $f = 1, 2, 3$, the subpacket labeling is 
\be
W^i_{\{f,\frac{1}{3}\}} = \left(W_{\{f,\frac{1}{3}\}}^{i,\{1,2\},1}, W_{\{f,\frac{1}{3}\}}^{i,\{1,2\},2}\right) \;\;\; : \;\;\; i = 1, \cdots, L.
\ee 
As a result, the caches are given by:
\be
Z_1 = \left\{W_{\{f,\frac{2}{3}\}}^{i,\{1\},1}, W_{\{f,\frac{1}{3}\}}^{i,\{1,2\},1}: i = 1, \cdots, L, f=1,2,3\right\}
\ee
\be
Z_2 = \left\{W_{\{f,\frac{2}{3}\}}^{i,\{2\},1}, W_{\{f,\frac{1}{3}\}}^{i,\{1,2\},1}: i = 1, \cdots, L, f=1,2,3\right\}
\ee
Assuming that, without loss of generality, user $1$ requests  packets $[s_1:s_1+L'-1]$ of file 
$W_1$ and users $2$ requests packets $[s_2:s_2+L'-1]$ of file  $W_2$, user $1$ sends $W_{\{2,\frac{2}{3}\}}^{i,\{1\},1}: i = s_2, \cdots, s_2+L'-1$ 
to user $2$ and user $2$ sends $W_{\{1,\frac{2}{3}\}}^{i,\{2\},1}: i = s_1, \cdots, s_1+L'-1$ to user $1$, such that the transmission rate 
is $R(2) = 1/3 \cdot 2 = \frac{2}{3}$.

It is also interesting to compute the converse (rate lower bound) for this case. In this case, for the sake of clarity, 
we use the same notation used in the example of Section \ref{sec:example} (see Fig.~\ref{fig: Network_Coding_Model}).
In particular, we label the three files as $A, B$ and $C$, and let $X_{u,\fsf}$ denote the codeword sent by user $u=1,2$ in the presence of 
the request vector $\fsf$.  Consider user $2$.  From the cut that separates $(X_{1,(A,B)}, X_{1,(B,C)}, X_{1,(C,A)}, Z_2)$ 
and $(\hat{W}_{2,A}, \hat{W}_{2,B}, \hat{W}_{2,C})$, by using the fact that the sum of the
entropies of the received messages and the entropy of the side information (cache symbols) cannot be
smaller than the number of requested information bits, we obtain that
\be
\label{eq: outter eg 1}
\sum_{s=1}^{\frac{L}{L'}} \left(R_{1,s, (A,B)}^{\rm T} + R_{1,s, (B,C)}^{\rm T} + R_{1,s,(C,A)}^{\rm T}\right) + MFL \geq 3FL.
\ee
Similarly, from the cut that separates $(X_{2,(A,B)}, X_{2,(B,C)}, X_{2,(C,A)}, Z_1)$ and $(\hat{W}_{1,A}, \hat{W}_{1,B}, \hat{W}_{1,C})$, we obtain 
\be
\label{eq: outter eg 2}
\sum_{s=1}^{\frac{L}{L'}} \left(R_{2,s,(A,B)}^{\rm T} + R_{2,s,(B,C)}^{\rm T} + R_{2,s, (C,A)}^{\rm T}\right) + MFL \geq 3FL.
\ee
By adding (\ref{eq: outter eg 1}) and (\ref{eq: outter eg 2}), we have
\begin{eqnarray}
&& \sum_{s=1}^{\frac{L}{L'}} \left(R_{1,s, (A,B)}^{\rm T} + R_{2,s,(A,B)}^{\rm T} + R_{1,s, (B,C)}^{\rm T} + R_{2,s,(B,C)}^{\rm T}  \right. \notag\\
&& \left. + R_{1,s,(C,A)}^{\rm T} + R_{2,s, (C,A)}^{\rm T} \right) + 2MFL
\geq 6FL.  \label{bimbominchia}
\end{eqnarray}
Since we are interested in minimizing the worst-case rate, the sum $R_{1,s, \fsf}^{\rm T} + R_{2,s,\fsf}^{\rm T}$ 
must yields the same min-max value $R^{\rm T}$ for any $s$ and $\fsf$. Then, (\ref{bimbominchia}) becomes
$3 R^{\rm T} + 2MFL' \geq 6FL'$, which yields
\be
R^{\rm T} \geq 2FL' - \frac{2FL'}{3}M.
\ee
Finally, dividing by $FL'$ we have
\be
R^*(M) = \frac{R^{\rm T}}{FL'} \geq 2-\frac{2}{3}M.
\ee
For $M = 2$, we have $R^*(2) \geq 2 - \frac{2}{3} \cdot 2 = \frac{2}{3}$, which shows the optimality of the cache sharing scheme in this case. 
\hfill $\lozenge$
\end{example}

\section{Proof of Theorem~\ref{theorem: 1} and Corollary~\ref{corollary: 1}}
\label{sec: Proof of Theorem 1 and Corollary 1}

With the caching placement and delivery schemes of Appendix \ref{Caching Placement and Delivery Scheme}
with integer $t = \frac{Mn}{m}$, any node in any subset of  $t+1$ nodes can transmit a subpacket that is useful for all the other $t$ nodes of the subset. 
Any subset of  $t+1$ nodes corresponds to $t+1$ (coded) transmissions, each of which has  block length $\frac{FL'}{t{n \choose t}}$ bits. 
In total, we have $(t+1){n \choose t+1}$ transmissions. Therefore, the total transmission length is
\begin{eqnarray}
\label{eq: R 1}
R^{\rm T} = (t+1){n \choose t+1} \cdot \frac{FL'}{t{n \choose t}}. 
\end{eqnarray}
Using $t = \frac{Mn}{m}$ in (\ref{eq: R 1}), we have
\begin{eqnarray}
R^{\rm T} = \left(\frac{n}{t} - 1\right) FL'
= \frac{m}{M}\left(1-\frac{M}{m}\right)FL'.
\end{eqnarray}
Finally, using the definition of rate, we have that the rate of the scheme is given by 
\be
\label{eq: t integer}
R(M) = \frac{R^{\rm T} }{FL'} = \frac{m}{M}\left(1-\frac{M}{m}\right).
\ee
When $t$ is not an integer, it is easy to see that the convex lower envelope of (\ref{eq: t integer}) is achievable 
(see Example \ref{example: 3} in Appendix \ref{Caching Placement and Delivery Scheme}).

\section{Proof of Theorem~\ref{theorem: 2}}
\label{Proof of Theorem 2}

In this section we generalize the cut-set bound method outlined in Section \ref{sec:example}.
Consider a two-dimensional augmented network layout of the type of Fig.~\ref{fig: Network_Coding_Model}, where a ``column'' of nodes 
corresponds to a user and a ``row'' of nodes corresponds to a demand vector $\fsf$ (for example, in  Fig.~\ref{fig: Network_Coding_Model}
the right-most column corresponds to user 1, and the top row corresponds to $\fsf = (A,B,C)$).
Directly by the problem definition, the cache message $Z_u$ is connected to all nodes of column $u$, and the 
coded (multicast) message $X_{u,\fsf}$ is connected to all nodes $v \neq u$ of row $\fsf$. 
In general, such graph has $n$ columns and $m^n$ rows. However, it is clear that by applying cut-set bound inequalities to 
the subgraph including only subset of such rows, i.e., a subset of the possible demand vectors, we obtain a lower bound to the best achievable 
rate $R^*(M)$. 

In particular, for the bound of Theorem~\ref{theorem: 2}, we need to consider two types of cuts. 
The first type includes $m$ requests vectors $\{\fsf_j : j = 1, \ldots, m\}$ (i.e., $m$ rows of the graph) constructed as follows. 
Consider the semi-infinite sequence periodic concatenation of the integers $[1,2, \ldots, m, 1,2, \ldots, m, 1, 2, \dots )$. Then, 
$\fsf_j$ is the vector of length $n$ formed by the components $[j: j+n-1]$ of such concatenation. For example, 
in the case $m < n$, the first few demand vectors in this set are
\begin{eqnarray*}
\fsf_1 &= &\{1, 2, 3, \cdots,  m-2, m-1, m, 1, 2, 3, \cdots, m-1, m, \cdots\}, \\
\fsf_2 &= & \{2, 3, 4, \cdots,  m-1, m, 1, 2, 3, 4, \cdots, m, 1, \cdots\}, \\
& \vdots & 
\end{eqnarray*}
while if  $m \geq n$ they are
\begin{eqnarray*}
\fsf_1 & = & \{1, 2, 3, \cdots, n-2, n-1, n \}, \\
\fsf_2 & = & \{2, 3, 4, \cdots, n-1, n, n+1\}, \\
 & \vdots & 
\end{eqnarray*}
Using the fact that the sum of the entropies of the received messages and the entropy of the side information (cache symbols) 
cannot be smaller than the number of requested information bits,
for each $v \in \Uc$, the cut-set bound applied to the cut that separates $\{Z_v, \{X_{u,\fsf_j}: j = 1, \cdots, m\} : \; \forall u \neq v\}$
and $\{\hat{W}_{u, f}: f = 1, \cdots, m\}$ yields
\be
\label{eq: outter general 1}
\sum_{s=1}^{\frac{L}{L'}} \sum_{u=1,u \neq v}^{n} \sum_{j=1}^m R_{u,s,\fsf_j}^{\rm T} + MFL \geq mFL. 
\ee
By summing (\ref{eq: outter general 1}) over all $v \in \Uc$, we obtain
\begin{eqnarray}
&&\sum_{v=1}^{n}\sum_{s=1}^{\frac{L}{L'}} \sum_{u=1,u \neq v}^{n} \sum_{j=1}^m R_{u,s,\fsf_j}^{\rm T} + nMFL 
\geq nmFL.
\end{eqnarray}
Since we are interested in minimizing the worst-case rate, the sum $\sum_{u=1}^nR_{u,s,\fsf_j}^{\rm T}$ must yields
the same min-max value $R^{\rm T}$ for any $s$ and $\fsf_j$. This yields the bound
\be
\label{eq: outter general 2}
\frac{L}{L'}(n-1)mR^{\rm T} + nMFL\geq nmFL.
\ee

Dividing both sides of (\ref{eq: outter general 2}) by $(n-1)mFL$ and using the definition of rate $R(M) = R^{\rm T}/(FL')$, we conclude that
the best possible achievable rate must satisfy
\be
R^*(M) \geq \frac{n}{n-1}\left(1-\frac{M}{m}\right),
\ee
which is the second term in the max in the right-hand side of (\ref{banana}). 
Notice that this bound provides the tight converse for Examples \ref{example: 1}, \ref{example: 2} and \ref{example: 3} 
in Appendix \ref{Caching Placement and Delivery Scheme}.

For the second type of cut, for $l=1,\cdots,\min\{m,n\}$, we consider the 
first $l$ users, with $\lfloor \frac{m}{l} \rfloor$ requests vectors\footnote{With some slight abuse of notation, 
here we focus only on the first $l$ components of the request vectors and yet we indicate these vectors by $\fsf_j$, 
meaning that the other $n - l$ elements are irrelevant for the bound.}
\be
\label{eq: request vectors}
\fsf_j = \{l(j-1)+1, \cdots, jl\}, \;\;\; j = 1, \cdots, \lfloor \frac{m}{l} \rfloor. 
\ee 
From the cut that separates $$\left \{\{Z_v : v = 1, \ldots, l\} , \{\{X_{u,\fsf_j} : j=1, \cdots, \lfloor \frac{m}{l} \rfloor\} \; \forall \; u \in \Uc \} \right \}$$ 
and $$\left \{\{ \hat{W}_{v,(j-1)l+v} :  j=1, \cdots, \lfloor \frac{m}{l} \rfloor\} : v = 1,\ldots, l  \right \},$$ 
we obtain the inequality
\begin{eqnarray}
\sum_{s=1}^{\frac{L}{L'}}\sum_{j=1}^{\left\lfloor\frac{m}{l}\right\rfloor} \sum_{u=1}^n R_{u, s, \fsf_j}^{\rm T}+ l MFL
\geq l\left\lfloor\frac{m}{l}\right\rfloor FL.
\end{eqnarray}
Since we are interested in minimizing the worst-case rate, the sum $\sum_{u=1}^n R_{u, s, \fsf_j}^{\rm T}$ must yields
the same min-max value $R^{\rm T}$ for any $s$ and $\fsf_j$. This yields the bound
\begin{eqnarray}
\frac{L}{L'}\left\lfloor\frac{m}{l}\right\rfloor R^{\rm T} + l MFL
\geq l\left\lfloor\frac{m}{l}\right\rfloor FL, 
\end{eqnarray}
which can be written as
\be
R^{\rm T} \geq \left(l - \frac{l}{\lfloor\frac{m}{l}\rfloor}M\right)FL'.
\ee
It follows that the optimal achievable rate must satisfy
\be
R^*(M) \geq \max_{l \in \{1, 2, \cdots, \min\{m, n\}\}} \left(l - \frac{l}{\lfloor\frac{m}{l}\rfloor}M\right),
\ee
which is the first term in the max in the right-hand side of (\ref{banana}).


\section{Proof of Theorem~\ref{theorem: gap 1} and Corollary \ref{corollary: gap naive multicast}}
\label{sec: proof of theorem 3}

We let $G = \frac{R(M)}{R^*(M)}$ denote the multiplicative gap between the rate achievable by our scheme and
the best possible achievable rate. Upper bounds on $G$ are obtained by bounding the ratio between the achievable rate 
our the proposed schemes and the converse lower bound of Theorem \ref{theorem: 2}. 
From  Theorem \ref{theorem: 1} we have
\be
\label{eq: achievable 1}
R(M) \leq \frac{n}{\lfloor t \rfloor} - 1.
\ee
Also, it is immediately evident that $\frac{t}{\lfloor t \rfloor} \leq 2$ for all $t \geq 1$. 
In order to prove Theorem~\ref{theorem: gap 1}, we distinguish between the cases $n = \omega(m)$  and $n = O(m)$. 

\subsection{Case $n = \omega(m)$}

In this case, by using (\ref{eq: achievable 1}), we have 
\begin{eqnarray}
\label{eq: achievable 2}
R(M) &\leq& \frac{n}{\lfloor \frac{nM}{m} \rfloor} - 1 \notag\\ 
&=& \frac{m}{M} - 1 + o\left(\frac{m}{M}\right). 
\end{eqnarray}

\subsubsection{When $\frac{1}{2} \leq M = o(m)$}

Let $l^* = \left\lfloor \frac{m}{2M} \right\rfloor$, then by using Theorem~\ref{theorem: 2}, we obtain
\begin{eqnarray}
\label{eq: converse sl 1}
R^*(M) &\geq& \left(l^* - \frac{l^*}{\left\lfloor \frac{m}{l^*} \right\rfloor} M\right) \notag\\
&=& \left(\left\lfloor \frac{m}{2M} \right\rfloor - \frac{\left\lfloor \frac{m}{2M} \right\rfloor}{\left\lfloor \frac{m}{\left\lfloor \frac{m}{2M} \right\rfloor} \right\rfloor} M\right) \notag\\
&=& \frac{m}{4M} + o\left(\frac{m}{4M}\right), 
\end{eqnarray}
so that we can write
\be
G  \leq \frac{\frac{m}{M} - 1 + o\left(\frac{m}{M}\right)}{\frac{m}{4M} + o\left(\frac{m}{4M}\right)} = 4 + o(1).
\ee

\subsubsection{When $M = \Theta(m)$}

\begin{itemize}
\item If $\frac{m}{2M} \geq 3$, let $l^* = \left\lfloor \frac{m}{2M} \right\rfloor$, then by using Theorem~\ref{theorem: 2}, we get
\begin{eqnarray}
\label{eq: converse sl 2}
R^*(M) 
&=& \left(\left\lfloor \frac{m}{2M} \right\rfloor - \frac{\left\lfloor \frac{m}{2M} \right\rfloor}{\left\lfloor \frac{m}{\left\lfloor \frac{m}{2M} \right\rfloor} \right\rfloor} M\right) \notag\\
&\geq& \left(\frac{m}{2M} - 1\right)\left(1 - \frac{M}{2M} + o(1)\right) \notag\\
&\geq& \frac{\frac{m}{2M} - 1}{2} + o(1), 
\end{eqnarray}
which yields
\be
\label{eq: gap sl 1}
G \leq \frac{\frac{m}{M} - 1 + o\left(\frac{m}{M}\right)}{\frac{\frac{m}{2M} - 1}{2} + o(1)} \leq \frac{2}{\frac{1}{2} - \frac{M}{m}} \leq 6 + o(1).
\ee
\item If $\frac{m}{2M} < 3$, let $l^* = 1$, by using Theorem~\ref{theorem: 2}, we obtain
\be
\label{eq: converse sl 3}
R^*(M) \geq 1-\frac{M}{m}.
\ee
Then, we have
\be
\label{eq: gap sl 2}
G \leq \frac{\frac{m}{M} - 1 + o\left(\frac{m}{M}\right)}{1-\frac{M}{m}} \leq \frac{m}{M} + o(1) \leq 6 + o(1). 
\ee
\end{itemize}

\subsubsection{When $M < \frac{1}{2}$}

Let $l^* = m$, by using Theorem~\ref{theorem: 2}, we have
\be
\label{eq: converse sl 8}
R^*(M) \geq m(1-M). 
\ee
Then, we obtain
\be
\label{eq: converse sl 7}
G \leq \frac{\frac{m}{M}-1+o(\frac{m}{M})}{m(1-M)} \leq \frac{1}{M(1-M)} + o(1) \leq \frac{2}{M} + o(1).
\ee

\subsection{Case $n=O(m)$}

\subsubsection{When $\frac{1}{2} \leq M = o(m)$}

By letting $l^* = \left\lfloor \frac{m}{2M} \right\rfloor$, the lower bound of $R^*(M)$ is given by (\ref{eq: converse sl 1}). 

\begin{itemize}

\item If $t = \frac{nM}{m} = \omega(1)$, the upper bound of $R(M)$ is given by (\ref{eq: achievable 2}).
Then, we obtain
\be
G \leq \frac{ \frac{m}{M} - 1 + o\left(\frac{m}{M}\right)}{\frac{m}{4M} + o\left(\frac{m}{M}\right)} = 4 + o(1).
\ee

\item If $t = \frac{nM}{m} = \Theta(1)$, 
by using (\ref{eq: achievable 1}), we have
\be
\label{eq: gap sl 3}
G \leq \frac{ \frac{n}{\lfloor t \rfloor}}{\frac{m}{4M} + o\left(\frac{m}{4M}\right)} = \frac{4t}{\lfloor t \rfloor} + o(1). 
\ee



\end{itemize}


\subsubsection{When $M = \Theta(m)$}

By using (\ref{eq: gap sl 1}) and (\ref{eq: gap sl 2}), we obtain
\be
G \leq 6 + o(1).
\ee

\subsubsection{When $M < \frac{1}{2}$}

\begin{itemize}

\item If $n \leq m$, by using (\ref{eq: achievable 1}), we have
\begin{eqnarray}
\label{eq: achievable 4}
R(M) \leq n. 
\end{eqnarray}
Let $l^* = n$, by using Theorem~\ref{theorem: 2}, we obtain
\begin{eqnarray}
\label{eq: converse sl 4}
R^*(M) &\geq& n\left(1 - \frac{M}{\left\lfloor \frac{m}{n} \right\rfloor}\right).
\end{eqnarray}
If $\frac{m}{n} \geq 2$, by using (\ref{eq: converse sl 4}), we have
\begin{eqnarray}
\label{eq: converse sl 5}
R^*(M) &\geq& n\left(1 - \frac{M}{\frac{m}{n} -1}\right) \notag\\
&\geq& n\left(1 - \frac{M}{2 -1}\right) \notag\\
&\geq& \frac{n}{2}.
\end{eqnarray}
If $\frac{m}{n} < 2$, by using (\ref{eq: converse sl 4}), we have
\begin{eqnarray}
\label{eq: converse sl 6}
R^*(M) &\geq& n\left(1 - M\right) \notag\\
&\geq& \frac{n}{2}.
\end{eqnarray}
Thus, by using (\ref{eq: achievable 4}), (\ref{eq: converse sl 5}) and (\ref{eq: converse sl 6}), we get
\be
G \leq \frac{n}{n/2} = 2.
\ee
\item If $n > m$, the lower bound of $R^*(M)$ is given by (\ref{eq: converse sl 8}). Then, we have 
\be
G \leq \frac{\frac{n}{\lfloor t \rfloor}}{m(1-M)} \leq \frac{t}{\lfloor t \rfloor} \frac{2}{M}.
\ee

\end{itemize}


The proof of Corollary \ref{corollary: gap naive multicast} follows the exact same step as in the proof of Theorem \ref{theorem: gap 1} 
with the exception of case $n > m$ and $M < \frac{1}{2}$. 
In this case, when $L'=L$ (i.e., naive multicast is allowed), 
then by multicasting all the requested subpackets
\begin{eqnarray}
\label{eq: achievable nm}
R(M) \leq m
\end{eqnarray}
is achievable. Hence, when $n > m$ and $M < \frac{1}{2}$, we let $l^* = m$ and, from Theorem \ref{theorem: 2} 
and (\ref{eq: achievable nm}), we have
\begin{eqnarray}
\label{eq: gap proof 5}
G \leq \frac{m}{(1-M)m} \leq \frac{m}{\frac{1}{2}m} \leq 2.
\end{eqnarray}

\section{Proof of Theorem~\ref{theorem: 3}}
\label{sec: Proof of Theorem 4}

In each cluster, by using Theorem~\ref{theorem: 1}, we have the total number of bits needed to be transmitted in each cluster is $\frac{m}{M}\left(1-\frac{M}{m}\right)FL'$, therefore, by Definition~\ref{def: Delivery Phase} and denoting the achievable rate for each cluster as $R_c(M)$, then when $t=\frac{g_cM}{m}$ is an integer, we have
\be
\label{eq: t gc intefer}
R_c(M) = \frac{m}{M}\left(1-\frac{M}{m}\right).
\ee
When $t$ is not an integer, then the convex lower envelope of (\ref{eq: t gc intefer}) is achievable. 
Hence, the achievable throughput is given by
\be
T(M) = \frac{C_r}{\Kc}\frac{1}{R_c(M)} = \frac{C_r}{\Kc}\frac{1}{R(M)}. 
\ee

\section{Proof of Theorem~\ref{theorem: 4}}
\label{sec: Proof of theorem 5}

Due to the protocol channel model, users have to be within a radius of $r$ to receive information simultaneously. Hence, the maximum number of users that can receive useful information simultaneously is $\pi r^2 n$.\footnote{Since we consider the asymptotic regime $n \rightarrow \infty$, we ignore the non-integer part of $\pi r^2 n$.} Similarly, only users within radius of $r$ of each of these $\pi r^2 n$ users can serve them. Therefore, the maximum number of users that can serve these $\pi r^2 n$ users 
is at most $4\pi r^2 n$. In this proof, we consider a particular group of served users within a radius of $r$ and with cardinality $\pi r^2 n$. For other users in the network, we assume they can be served by some genies without any cost. 

We first compute a lower bound of the min-max number of bits $R^{\rm T}$ needed to serve these $\pi r^2 n$ users. 
Similar as (\ref{eq: request vectors}), 
we consider the first $l$ users in the network, with $\lfloor \frac{m}{l} \rfloor$ requests vectors: 
\be
\label{eq: request vectors 2}
\fsf = \fsf_j = \{l(j-1)+1, \cdots, jl\},
\ee 
where $j = 1, \cdots, \lfloor \frac{m}{l} \rfloor$ and $l=1,\cdots,\min\{m,\pi r^2 n\}$. 
from the cut that separates $\{(Z_v, X_{u,\fsf_j}): v=1,\cdots, l, u = 1, \cdots, \lceil4\pi r^2 n\rceil, j=1, \cdots, \lfloor \frac{m}{l} \rfloor\}$ and $\{\hat{W}_{l,f}:  f=1, \cdots, l\lfloor \frac{m}{l} \rfloor\}$ and by using the fact that the sum of the entropies of the received messages and the entropy of the side information (cache symbols) cannot be smaller than the number of requested information bits, we obtain that 
\begin{eqnarray}
\sum_{s=1}^{\frac{L}{L'}}\sum_{j=1}^{\left\lfloor\frac{m}{l}\right\rfloor} \sum_{u=1}^{\lceil4\pi r^2 n\rceil} R_{u, s, \fsf_j}^{\rm T}+ l MFL
& \geq & \left\lfloor\frac{m}{l}\right\rfloor\frac{L}{L'} \cdot R^{\rm T} + l MFL \notag\\
& \geq & l\left\lfloor\frac{m}{l}\right\rfloor FL.
\end{eqnarray}

Then, we can obtain
\be
R^{\rm T} \geq \left(l - \frac{l}{\left\lfloor\frac{m}{l}\right\rfloor}M\right)FL'.
\ee
Hence, for these served  $\pi r^2 n$ users, 
\be
R^*(M) \geq \max_{l \in \{1, 2, \cdots, \min\{m, \pi r^2 n\}\}}\left(l - \frac{l}{\left\lfloor\frac{m}{l}\right\rfloor}M\right).
\ee


We notice that it is possible to have multiple concurrent transmissions to serve these $\pi r^2 n$ users. The argument used here is similar as the one used in \cite{ji2013optimalJ}. By using the protocol model, since each node consumes at most the area of a disk with radius $(1+\Delta)r$, we can see that the total area consumed by all the nodes in a disk with radius $r$ is at most $\pi (r+(1+\Delta)r)^2 = (2+\Delta)^2\pi r^2$. 
Since each transmission consumes at least the area of a disk with radius of $\frac{\Delta}{2}r$ (See the proof of Theorem 1 in \cite{ji2013optimalJ} for details), we can obtain the maximum number of concurrent transmissions is 
\be
\left\lceil \frac{(2+\Delta)^2\pi r^2}{\pi \left(\frac{\Delta}{2}r\right)^2} \right\rceil  = \left\lceil \frac{4(2+\Delta)^2}{\Delta^2} \right\rceil.
\ee
Thus, there are at most $\left\lceil \frac{4(2+\Delta)^2}{\Delta^2} \right\rceil$ concurrent transmissions to serve the users in the disk with radius $r$. Therefore, the achievable throughput is given by
\begin{eqnarray}
T(M) & = & \left\lceil \frac{4(2+\Delta)^2}{\Delta^2} \right\rceil \frac{C_r}{R(M)} \notag\\
& \leq & C_r \left\lceil \frac{4(2+\Delta)^2}{\Delta^2} \right\rceil \frac{1}{\max_{l \in \{1, 2, \cdots, \min\{m, \pi r^2 n\}\}}\left(l - \frac{l}{\left\lfloor\frac{m}{l}\right\rfloor}M\right)}.
\end{eqnarray}

\section{Proof of Theorem \ref{theorem: gap 2}}
\label{sec: Proof of theorem gap}

By using Theorem \ref{theorem: 3} and Theorem \ref{theorem: 4}, we obtain
\begin{eqnarray}
\frac{T^*(M)}{T(M)} & \leq & \frac{C_r \left\lceil \frac{4(2+\Delta)^2}{\Delta^2} \right\rceil \frac{1}{\max_{l \in \{1, 2, \cdots, \min\{m, \pi r^2 n\}\}}\left(l - \frac{l}{\left\lfloor\frac{m}{l}\right\rfloor}M\right)}}{\frac{C_r}{\Kc}\frac{1}{R(M)}} \notag\\
& = & \Kc \left\lceil \frac{4(2+\Delta)^2}{\Delta^2} \right\rceil \frac{R(M)}{\max_{l \in \{1, 2, \cdots, \min\{m, \pi r^2 n\}\}}\left(l - \frac{l}{\left\lfloor\frac{m}{l}\right\rfloor}M\right)} \notag\\
&\buildrel (a) \over \leq & \Kc \left\lceil \frac{4(2+\Delta)^2}{\Delta^2} \right\rceil \times \left\{\begin{array}{cc}4, &   t = \omega(1), \frac{1}{2} \leq M = o(m) \\ 
\frac{4t}{\lfloor t \rfloor}, &  \pi r^2 n  = O(m), t = \Theta(1), \frac{1}{2} \leq M = o(m) \\
6, & M = \Theta(m) \\
\frac{2}{M}, &  \pi r^2 n  = \omega(m), M <\frac{1}{2} \\
\frac{t}{\lfloor t \rfloor} \frac{2}{M}, &n = O(m),  \pi r^2 n  > m, M <\frac{1}{2}\\
2, & n = O(m),  \pi r^2 n  \leq m, M < \frac{1}{2} 
\end{array}\right.,
\end{eqnarray}
where (a) is obtained by Theorem \ref{theorem: gap 1}, in which $n$ is replaced by $\pi r^2 n$.

\section{Proof of Theorem \ref{theorem: caching}}
\label{sec: Proof of theorem caching}

We need to determine the value of $\rho$ such that the network can cache at least $K$ distinct MDS-coded symbols of each packet from each file with 
high probability as $K \rightarrow \infty$ by using Algorithm~\ref{algorithm: 5}. For the sake of analysis, 
we consider a simpler algorithm whose performance is worse than Algorithm \ref{algorithm: 5}, but it turns out to be good enough
to prove our result.  In this new algorithm, each user selects $MK/m$ MDS-coded symbols 
independently with uniform probability (with probability $\frac{\rho}{K}$) of a packet from each file. 
Selection is done with replacement, i.e., there is the possibility of choosing the same coded symbol multiple times.
Hence, this simplified selection method is certainly not better than the selection in Algorithm~\ref{algorithm: 5} (selection without replacement). 

As in Algorithm~\ref{algorithm: 5}, each user caches the same set of MDS-coded symbols of each packet from each file. 
Hence, in the following, we refer to ``MDS-coded symbol'' $i$ without specifying which file and which packet in the file it belongs to, 
since this will be the same for all files and all packets in each file.
We denote by $Z$ the number of distinct MDS-coded symbols (same for each packet from each file) 
obtained by the new algorithm. Notice that  $Z = \sum_{i=1}^{\frac{K}{\rho}}I_i$, where $I_i = 1$ if MDS-coded symbol $i$ 
is cached, otherwise $I_i = 0$. 
Similarly, let $Z' = \sum_{i=1}^{\frac{K}{\rho}}Y_i$ be the number of distinct MDS-coded symbols 
obtained by using Algorithm \ref{algorithm: 5}, where $Y_i$ is an indicator function similarly defined as $I_i$. 
Notice that Algorithm \ref{algorithm: 5} stochastically dominates the simplified algorithm 
in terms of the number of distinct MDS-coded symbols of a packet from each file, i.e., for any $a > 0$ we have
\be
\label{eq: algorithm 5}
\PP(Z' \leq a) \leq \PP(Z \leq a).
\ee
In particular, if we show that $\PP(Z > K) \rightarrow 1$ as $K \rightarrow \infty$, the stochastic dominance
(\ref{eq: algorithm 5}) immediately implies that also $\PP(Z'  > K) \rightarrow 1$ as $K \rightarrow \infty$. 

Noticing that we have $n$ users, each of which
makes $MK/m$ independent selections with uniform probability over a set of $K/\rho$ possible MDS-coded symbol indices,
we have that $I_i$ is Bernoulli with probability $\PP(I_i = 0) =  (1 - \frac{\rho}{K})^{nMK/m}$. This yields
\be
\EE[Z] = \frac{K}{\rho}\left(1 - \left(1-\frac{\rho}{K}\right)^{\frac{nMK}{m}}\right).
\ee
Then, we have
\begin{eqnarray}  \label{lbEZ}
\EE\left[Z\right] 
 \geq \frac{1 - \exp(-t \rho)}{\rho} K,
\end{eqnarray}
where $t = \frac{Mn}{m} > 1$.  
The proof of Theorem \ref{theorem: caching} is obtained in the following steps.
First, we consider the concentration of $Z$ around its mean $\EE[Z]$. Then, we find
that for $t > 1$ it is possible to find $\rho \in (0,1)$ such that 
$\EE[Z] = (1 + \delta) K$, where $\delta > 0$ is a constant independent of $K$. 
Combining these results, we have that, as $K \rightarrow \infty$, 
the number of cached MDS-coded symbols in the network for all packets of all files is larger than $K$ with
probability growing to 1. 

We start by considering the concentration of $Z$. 
To this purpose, we recall here the definition of self-bounding function \cite{boucheron2013concentration}:
\begin{defn}
\label{def: selfbounding}
A nonnegative function $f: \mathcal{X}^n \rightarrow [0, \infty)$ has the self-bounding property 
if there exist functions $f_i: \mathcal{X}^{n-1} \rightarrow \mathbb{R}$ such that for all $x_1, \cdots, x_n \in \mathcal{X}$ and all $i=1, \cdots, n$,
\be
\label{eq: selfbounding 1}
0 \leq f(x_1, \cdots, x_n) - f_i(x_1, \cdots, x_{i-1}, x_{i+1}, \cdots, x_n) \leq 1,
\ee
and also
\be
\label{eq: selfbounding 2}
\sum_{i=1}^n\left( f(x_1, \cdots, x_n) - f_i(x_1, \cdots, x_{i-1}, x_{i+1}, \cdots, x_n) \right) \leq f(x_1, \cdots, x_n).
\ee
\hfill $\lozenge$
\end{defn}
We observe that $Z$ is a self-bounding function of the $I_i$'s. To see this, 
let $n = K/\rho$, $x_i = I_i$ for $i = 1,\ldots, K/\rho$, $Z = f(I_1,\ldots,I_{K/\rho}) = \sum_{j=1}^{K/\rho} I_j$ and $Z_i = f_i(I_1, \ldots, I_{K/\rho}) = 
\sum_{j\neq i} I_j$.  Then, $Z - Z_i = I_i \in \{0,1\}$ such that (\ref{eq: selfbounding 1}) holds. Furthermore, 
$\sum_{i=1}^{K/\rho} (Z - Z_i) = \sum_{i=1}^{K/\rho} I_i = Z$ such that also (\ref{eq: selfbounding 2}) holds. 
As a consequence, we have \cite{boucheron2013concentration}:
\begin{lemma}
\label{lemma: exponential bound}
If $Z$ has the self-bounding property, then for every $0 < \mu \leq \EE[Z]$,
\be
\label{eq: exponential bound}
\PP(Z - \EE[Z] \geq \mu) \leq \exp\left(-h\left(\frac{\mu}{\EE[Z]}\right) \EE[Z]\right),
\ee
and
\be
\label{eq: exponential bound 2}
\PP(Z - \EE[Z] \leq -\mu) \leq \exp\left(-h\left(-\frac{\mu}{\EE[Z]}\right) \EE[Z]\right),
\ee
where $h(u) = (1+u)\log(1+u) - u$, $u \geq -1$.
\hfill  $\square$
\end{lemma}
Next, we are interested in studying the 
the quantities $h\left(\frac{\mu}{\EE[Z]}\right)$ and $h\left(-\frac{\mu}{\EE[Z]}\right)$ 
in the case $\mu = o\left(\EE[Z]\right)$.  We have
\begin{eqnarray}
h\left(\frac{\mu}{\EE[Z]}\right) &=& \left(1+\frac{\mu}{\EE[Z]}\right)\log\left(1+ \frac{\mu}{\EE[Z]}\right) - \frac{\mu}{\EE[Z]} \notag\\
& = & \left(1+\frac{\mu}{\EE[Z]}\right) \left(\frac{\mu}{\EE[Z]} + o\left(\left(\frac{\mu}{\EE[Z]}\right)^2\right)\right) - \frac{\mu}{\EE[Z]} \notag\\
& = & \frac{\mu}{\EE[Z]} + \frac{\mu^2}{\EE[Z]^2} - \frac{\mu}{\EE[Z]} + o\left(\left(\frac{\mu}{\EE[Z]}\right)^2\right) \notag\\
& = & \frac{\mu^2}{\EE[Z]^2} + o\left(\left(\frac{\mu}{\EE[Z]}\right)^2\right),
\end{eqnarray}
and
\begin{eqnarray}
h\left(-\frac{\mu}{\EE[Z]}\right) &=& \left(1-\frac{\mu}{\EE[Z]}\right)\log\left(1- \frac{\mu}{\EE[Z]}\right) + \frac{\mu}{\EE[Z]} \notag\\
& = & \left(1-\frac{\mu}{\EE[Z]}\right) \left(- \frac{\mu}{\EE[Z]} + o\left(\left(\frac{\mu}{\EE[Z]}\right)^2\right)\right) + \frac{\mu}{\EE[Z]} \notag\\
& = & -\frac{\mu}{\EE[Z]} + \frac{\mu^2}{\EE[Z]^2}+ \frac{\mu}{\EE[Z]} + o\left(\left(\frac{\mu}{\EE[Z]}\right)^2\right) \notag\\
& = & \frac{\mu^2}{\EE[Z]^2} + o\left(\left(\frac{\mu}{\EE[Z]}\right)^2\right).
\end{eqnarray}
Using the above results in (\ref{eq: exponential bound}) and in (\ref{eq: exponential bound 2}), and applying the union bound, we have
\begin{equation} \label{boundPP}
\PP(|Z - \EE[Z]| \geq \mu) \leq 2 \exp\left(-\frac{\mu^2}{\EE[Z]} + o\left(\frac{\mu^2}{\EE[Z]}\right) \right).
\end{equation} 
For what said above, Theorem \ref{theorem: caching} is proved if we find $\rho \in (0,1)$ such that
$\EE[Z] > (1+\delta)K$ for some $\delta > 0$ independent of $K$, and find $\mu$ such that
$\mu/\EE[Z] \rightarrow 0$ and $\mu^2 / \EE[Z] \rightarrow \infty$ as $K \rightarrow \infty$. 

To this purpose, we have:
\begin{lemma}
\label{lemma: fixed point}
For $t > 1$, the equation:
\be
\label{eq: fixed point 1}
x = 1 - \exp(-t x) 
\ee
has a unique solution $\rho^* \in (0,1)$. Furthermore,  
$\frac{1 - \exp(-t \rho)}{\rho} > 1$ for $0 < \rho < \rho^*$. 
\hfill  $\square$
\end{lemma}
\begin{IEEEproof}
Consider the function $f(x) = 1 - \exp(-xt)$, with derivatives
\[ f'(x) = t \exp(-tx), \;\; f''(x) = -t^2 \exp(-xt). \]
This is a monotonically increasing concave function, with slope at $x = 0$ equal to $t > 1$, and a horizontal asymptote
$\lim_{x \rightarrow \infty} f(x) = 1$. 
Since $f(0)= 0 $ and the slope at the origin is larger than 1, we have that $f(x) > x$ in a right neighborhood of $x = 0$. 
Since the slope for large $x$ is smaller than 1, we have that $f(x) < x$ for sufficiently large $x$. 
Hence, since $f(x)$ is continuous, $f(x) = x$ must have a strictly positive solution $x = \rho^* < 1$. 
Furthermore, given the concavity and monotonicity,  this solution must be unique, such that 
$f(x) > x$ for $x \in (0,\rho^*)$ and $f(x) < x$ for $x \in (\rho^*, +\infty)$. This also implies that 
the iteration $x^{(\ell)} = f(x^{(\ell-1)})$ for $\ell = 1,2,3,\ldots$ yields a monotonically increasing sequence 
uniformly upper bounded by $\rho^*$ for any initial condition $x^{(0)} \in (0,\rho^*)$ and a monotonically decreasing sequence 
uniformly lower bounded by $\rho^*$ for all initial conditions $x^{(0)} \in (\rho^*,+\infty)$. It is immediate to see that both these sequences 
converge to $\rho^*$, otherwise this would contradict the uniqueness of the strictly positive solution of $x = f(x)$. 
Finally, for any $0 < \rho < \rho^*$, $f(\rho) > \rho$ implies $f(\rho)/\rho > 1$.
\end{IEEEproof}

Letting $\rho^*$ denote the unique positive solution of (\ref{eq: fixed point 1}), we choose $\rho$ such that 
\be
\rho = (1 - \varepsilon) \rho ^*, 
\ee
where $\varepsilon > 0$ is small enough such that $\rho > 0$. Then, Lemma \ref{lemma: fixed point} and  the lower bound (\ref{lbEZ}) imply
\[ \EE[Z] = (1+\delta(\varepsilon))K \]
for some $\delta(\varepsilon) > 0$ that does not depend on $K$. 

Letting $\mu = (1+\delta(\varepsilon))^{\frac{1}{2}}K^{\frac{1}{2} + \frac{\delta_1}{2}}$ 
for some constant $\delta_1 > 0$ and using (\ref{boundPP}), we obtain
\begin{eqnarray}
\PP(|Z - (1+\delta(\varepsilon))K| \geq \mu) & \leq & 2 \exp\left(-\frac{\left((1+\delta(\varepsilon))^{\frac{1}{2}}K^{\frac{1}{2} + \frac{\delta_1}{2}}\right)^2}{(1+\delta(\varepsilon))K} + o\left(\frac{\left(K^{\frac{1}{2} + \frac{\delta_1}{2}}\right)^2}{(1+\delta(\varepsilon))K} \right) \right) \notag\\
&=& \exp\left(-K^{\delta_1} + o\left(K^{\delta_1} \right)\right).
\end{eqnarray}
Thus, using the stochastic dominance (\ref{eq: algorithm 5}) we have immediately that, for $K \rightarrow \infty$,  
\be
\PP(Z' \geq K) \geq 1 -  \exp\left(-K^{\delta_1} + o\left(K^{\delta_1}\right)\right), 
\ee
and Theorem \ref{theorem: caching} is proved.

\section{Proof of Theorem \ref{theorem: decentralized 1} and of Lemma \ref{upper bound decentralized}}
\label{sec: Proof of Theorem theorem: decentralized 1}

First, we show the first term in (\ref{eq: decentralized 1}). By using the caching placement scheme given in {Algorithm}~\ref{algorithm: 5}, we can see that the probability that each MDS-coded symbol is stored in each node is given by
\begin{eqnarray}
\PP(\text{Each MDS-coded symbol is stored in each node}) & = & \frac{{\frac{K}{\rho}-1 \choose \frac{KM}{m}-1}}{{\frac{K}{\rho} \choose \frac{KM}{m}}} \notag\\
& = & \frac{(\frac{K}{\rho}-1)!}{(\frac{K}{\rho}-\frac{KM}{m})!\left(\frac{KM}{m}-1\right)!}\frac{(\frac{K}{\rho}-\frac{KM}{m})!\left(\frac{KM}{m}\right)!}{\frac{K}{\rho}!} \notag\\
& = & \frac{M\rho}{m}.
\end{eqnarray}

The expected number of MDS-coded symbols of each packet from each file that are cached exclusively at particular $s$ users is given by
\be
\frac{K}{\rho}\left(\frac{M\rho}{m}\right)^s\left(1-\frac{M\rho}{m}\right)^{n-s},
\ee
and when $K$ goes to infinity, then the actual number of MDS-coded symbol of each packet from each file that are cached exclusively at particular $s$ users is given by
\be
\frac{K}{\rho}\left(\frac{M\rho}{m}\right)^s\left(1-\frac{M\rho}{m}\right)^{n-s} + o\left(\frac{K}{\rho}\left(\frac{M\rho}{m}\right)^s\left(1-\frac{M\rho}{m}\right)^{n-s}\right)
\ee

Then, we have
\begin{eqnarray}
R^{\rm T} & = & \frac{K}{\rho}\sum_{s=2}^n s{n \choose s}\frac{1}{s-1}\left(\frac{M\rho}{m}\right)^{s-1}\left(1-\frac{M\rho}{m}\right)^{n-s+1} \frac{FL'}{K} \label{eq: RT D2D 2} \\
& = & \frac{K}{\rho}\sum_{s=2}^n\left(1+\frac{1}{s-1}\right){n \choose s}\left(\frac{M\rho}{m}\right)^{s-1}\left(1-\frac{M\rho}{m}\right)^{n-s+1} \frac{FL'}{K}\notag\\
& = & \frac{K}{\rho}\left(\sum_{s=2}^n{n \choose s}\left(\frac{M\rho}{m}\right)^{s-1}\left(1-\frac{M\rho}{m}\right)^{n-s+1}\right) \frac{FL'}{K} \notag\\
& & + \frac{K}{\rho}\sum_{s=2}^n\frac{1}{s-1}{n \choose s}\left(\frac{M\rho}{m}\right)^{s-1}\left(1-\frac{M\rho}{m}\right)^{n-s+1} \frac{FL'}{K}\notag\\
& = & \frac{K}{\rho}\frac{1-\frac{M\rho}{m}}{\frac{M\rho}{m}}\left(\sum_{s=2}^n{n \choose s}\left(\frac{M\rho}{m}\right)^{s}\left(1-\frac{M\rho}{m}\right)^{n-s}\right) \frac{FL'}{K} \notag\\
& & + \frac{K}{\rho}\sum_{s=2}^n\frac{1}{s-1}{n \choose s}\left(\frac{M\rho}{m}\right)^{s-1}\left(1-\frac{M\rho}{m}\right)^{n-s+1}\frac{FL'}{K}. \label{eq: RT D2D 1}
\end{eqnarray}
We divide $R^{\rm T}$ in (\ref{eq: RT D2D 2}) by $FL'$ and obtain the first term of (\ref{eq: decentralized 1}). 

Next,  we wish to show the first term of (\ref{eq: decentralized 2}) in Lemma \ref{upper bound decentralized}.
The first term of (\ref{eq: RT D2D 1}) can be computed as
\begin{eqnarray}
\label{eq: RT D2D 11}
& & \frac{K}{\rho}\frac{1-\frac{M\rho}{m}}{\frac{M\rho}{m}}\left(\sum_{s=2}^n{n \choose s}\left(\frac{M\rho}{m}\right)^{s}\left(1-\frac{M\rho}{m}\right)^{n-s}\right) \frac{FL'}{K} \notag\\
& & = \frac{K}{\rho}\frac{m}{M\rho}\left(1-\frac{M\rho}{m}\right)\left(1 - \left(1-\frac{M\rho}{m}\right)^n - \frac{M\rho n}{m}\left(1-\frac{M\rho}{m}\right)^{n-1}\right) \frac{FL'}{K}.
\end{eqnarray}

The second term of (\ref{eq: RT D2D 1}) is given by
\begin{eqnarray}
\label{eq: RT D2D 12}
& & \frac{K}{\rho}\sum_{s=2}^n\frac{1}{s-1}{n \choose s}\left(\frac{M\rho}{m}\right)^{s-1}\left(1-\frac{M\rho}{m}\right)^{n-s+1} \frac{FL'}{K} \notag\\
& & = \frac{K}{\rho}\frac{m}{MA}\left(1-\frac{M\rho}{m}\right)\sum_{s=2}^n\frac{1}{s-1}{n \choose s}\left(\frac{M\rho}{m}\right)^{s}\left(1-\frac{M\rho}{m}\right)^{n-s} \frac{FL'}{K} \notag\\
& & \leq \frac{K}{\rho}\frac{m}{M\rho}\left(1-\frac{M\rho}{m}\right)\sum_{s=2}^n\frac{1+2}{s+1}{n \choose s}\left(\frac{M\rho}{m}\right)^{s}\left(1-\frac{M\rho}{m}\right)^{n-s} \frac{FL'}{K} \notag\\
& & = \frac{3K}{\rho}\frac{m}{M\rho}\left(1-\frac{M\rho}{m}\right)\sum_{s=2}^n\frac{1}{s+1}{n \choose s}\left(\frac{M\rho}{m}\right)^{s}\left(1-\frac{M\rho}{m}\right)^{n-s} \frac{FL'}{K} \notag\\
& & = \frac{3K}{\rho}\frac{m}{M\rho}\left(1-\frac{M\rho}{m}\right) \left(\EE\left[\frac{1}{S+1}\right] - \left(1-\frac{M\rho}{m}\right)^n-\frac{1}{2}\frac{M\rho n}{m}\left(1-\frac{M\rho}{m}\right)^{n-1}\right) \frac{FL'}{K}, 
\end{eqnarray}
where $S$ is a random variable with Binomial distribution with parameters $n$ and $p=\frac{M\rho}{m}$.
Then,  we can compute $\EE\left[\frac{1}{S+1}\right]$ as
\begin{eqnarray}
\label{eq: EE[1/s+1]}
\EE\left[\frac{1}{S+1}\right] & = & \sum_{i = 0}^n \frac{1}{i+1} \frac{n!}{i!(n-i)!}p^i(1-p)^{n-i} \notag\\
& = & \sum_{i = 0}^n \frac{n!}{(i+1)!(n-i)!}p^i(1-p)^{n-i} \notag\\
& = & \sum_{i = 0}^n \frac{(n+1)!}{(i+1)!(n+1-i-1)!}\frac{1}{n+1}p^i(1-p)^{n-i} \notag\\
& = & \sum_{i = 0}^n {n+1 \choose i+1} \frac{1}{n+1}p^i(1-p)^{n-i} \notag\\
& = & \frac{1}{(n+1)p} \sum_{i = 0}^n{n+1 \choose i+1}p^{i+1}(1-p)^{n+1-i-1} \notag\\
& = & \frac{1}{(n+1)p} (1-(1-p)^{n+1}) \notag\\
& = & \frac{1}{(n+1)\frac{M\rho}{m}} \left(1-\left(1-\frac{M\rho}{m}\right)^{n+1}\right). 
\end{eqnarray}

Thus, plugging (\ref{eq: EE[1/s+1]}) into (\ref{eq: RT D2D 12}), we have
\begin{eqnarray}
\label{eq: RT D2D 121}
& & \frac{K}{\rho}\sum_{s=2}^n\frac{1}{s-1}{n \choose s}\left(\frac{M\rho}{m}\right)^{s-1}\left(1-\frac{M\rho}{m}\right)^{n-s+1} \frac{FL'}{K} \notag\\
& & = \frac{3K}{\rho}\frac{m}{M\rho}\left(1-\frac{M\rho}{m}\right) \left(\EE\left[\frac{1}{s+1}\right] - \left(1-\frac{M\rho}{m}\right)^n-\frac{1}{2}\frac{M\rho n}{m}\left(1-\frac{M\rho}{m}\right)^{n-1}\right) \frac{FL'}{K} \notag\\
& & = \frac{3K}{\rho}\frac{m}{M\rho}\left(1-\frac{M\rho}{m}\right)  \notag\\
& & \cdot \left(\frac{1}{(n+1)\frac{M\rho}{m}} \left(1-\left(1-\frac{M\rho}{m}\right)^{n+1}\right) - \left(1-\frac{M\rho}{m}\right)^n-\frac{1}{2}\frac{M\rho n}{m}\left(1-\frac{M\rho}{m}\right)^{n-1}\right) \frac{FL'}{K}. \notag\\
\end{eqnarray}
Then, using (\ref{eq: RT D2D 11}) and (\ref{eq: RT D2D 121}) into (\ref{eq: RT D2D 1}),  we obtain
\begin{eqnarray}
\label{eq: RT D2D 3}
R^{\rm T} & \leq &  \frac{K}{\rho}\frac{m}{M\rho}\left(1-\frac{M\rho}{m}\right)\left(1 - \left(1-\frac{M\rho}{m}\right)^n - \frac{M\rho n}{m}\left(1-\frac{M\rho}{m}\right)^{n-1}\right) \frac{FL'}{K} \notag\\
& & +  \frac{3K}{\rho}\frac{m}{M\rho}\left(1-\frac{M\rho}{m}\right) \left(\frac{1}{(n+1)\frac{M\rho}{m}} \left(1-\left(1-\frac{M\rho}{m}\right)^{n+1}\right) \right. \notag\\
& & \left. - \left(1-\frac{M\rho}{m}\right)^n-\frac{1}{2}\frac{M\rho n}{m}\left(1-\frac{M\rho}{m}\right)^{n-1}\right) \frac{FL'}{K} \notag\\
& = & \frac{K}{\rho}\frac{m}{M\rho}\left(1-\frac{M\rho}{m}\right)\left(1 - \left(1-\frac{M\rho}{m}\right)^n - \frac{M\rho n}{m}\left(1-\frac{M\rho}{m}\right)^{n-1} \right. \notag\\
& & \left. + \frac{3}{(n+1)\frac{M\rho}{m}} \left(1-\left(1-\frac{M\rho}{m}\right)^{n+1}\right) - 3\left(1-\frac{M\rho}{m}\right)^n-\frac{3}{2}\frac{M\rho n}{m}\left(1-\frac{M\rho}{m}\right)^{n-1}\right) \frac{FL'}{K}. \notag\\
& = & \frac{K}{\rho}\frac{m}{M\rho}\left(1-\frac{M\rho}{m}\right) \notag\\
&& \left(1+\frac{3}{(n+1)\frac{M\rho}{m}} \left(1-\left(1-\frac{M\rho}{m}\right)^{n+1}\right)
 - 4\left(1-\frac{M\rho}{m}\right)^n - \frac{5}{2}\frac{M\rho n}{m}\left(1-\frac{M\rho}{m}\right)^{n-1}\right)\frac{FL'}{K}. \notag\\
\end{eqnarray}
Therefore, dividing both sides of (\ref{eq: RT D2D 3}) by $FL'$, we have
\begin{eqnarray}
R(M) & = & \frac{R^{\rm T}}{FL'} \notag\\
&\leq& \frac{m}{M\rho^2}\left(1-\frac{M\rho}{m}\right) \notag\\
&& \left(1+\frac{3}{(n+1)\frac{M\rho}{m}} \left(1-\left(1-\frac{M\rho}{m}\right)^{n+1}\right)
 - 4\left(1-\frac{M\rho}{m}\right)^n - \frac{5}{2}\frac{M\rho n}{m}\left(1-\frac{M\rho}{m}\right)^{n-1}\right). \notag\\
\end{eqnarray}

The second term of (\ref{eq: decentralized 1}) and (\ref{eq: decentralized 2}) is obtained by counting the needed codewords (i.e., blocks of
linear hashed symbols)  such that all the users can successfully decode. It can be seen that each user need $(1-\frac{MK}{m})\frac{FL'}{K}$ 
codewords to decode. Hence, the total number of codewords needed to be transmitted is $(1-\frac{MK}{m})\frac{FL'}{K} \cdot n$. Therefore, we have  
\be
R(M) = \frac{R^{\rm T}}{FL'} = \frac{(1-\frac{MK}{m})\frac{FL'}{K} \cdot n}{FL'} = n-t.
\ee

%

%
%

\section{Proof of Theorem \ref{theorem: decentralized gap}}
\label{sec: proof theorem decentralized gap}

\subsection{Case $t = \omega(1)$}
By using Theorem \ref{theorem: decentralized 1}, we obtain
\begin{eqnarray}
\label{eq: achievable dc 1}
R(M) & = & \frac{m}{M\rho^2}\left(1-\frac{M\rho}{m}\right) \notag\\
&& \left(1+\frac{3}{(n+1)\frac{M\rho}{m}} \left(1-\left(1-\frac{M\rho}{m}\right)^{n+1}\right)
 - 4\left(1-\frac{M\rho}{m}\right)^n - \frac{5}{2}\frac{M\rho n}{m}\left(1-\frac{M\rho}{m}\right)^{n-1}\right) \notag\\
&=& \frac{m}{M\rho^2}\left(1-\frac{M\rho}{m}\right) \left(1 + \frac{3}{\rho t}\left(1-e^{-\rho t}\right) - 4 e^{-\rho t} - \frac{5}{2}\rho t e^{-\rho t} + o(1)\right) \notag\\
&=& \frac{m}{M\rho^2}\left(1-\frac{M\rho}{m}\right)\left(1 + \frac{3}{\rho t} - e^{-\rho t}\left(\frac{3}{\rho t} + 4 + \frac{5}{2}\rho t\right) + o(1)\right) \notag\\
&=& \frac{m}{M\rho^2}\left(1-\frac{M\rho}{m}\right)\left(1 + f_\rho(t)\right) \notag\\
&=& \frac{m}{M}\left(1-\frac{M}{m}\right) \frac{1-\frac{M\rho}{m}}{1-\frac{M}{m}} \frac{1}{\rho^2}  \left(1 + f_\rho(t)\right),
\end{eqnarray}
where $f_\rho(t) = \frac{3}{\rho t} - e^{-\rho t}\left(\frac{3}{\rho t} + 4 + \frac{5}{2}\rho t\right)$ is a function of $t$.\footnote{Notice that $\rho$ is a function of $t$.} 

If $t \rightarrow \infty$, by using (\ref{eq: achievable dc 1}), let $\rho = (1-\varepsilon)\rho^*$ and $M \leq \frac{m}{1+\varepsilon}$, where $\varepsilon$ is an arbitrary small positive number and $\rho^*$ is given by Theorem \ref{theorem: caching}, we obtain $f_\rho(t) \rightarrow 0$, $\rho \rightarrow 1-\varepsilon$ and
\begin{eqnarray}
R(M) &\leq& \frac{m}{M}\left(1-\frac{M}{m}\right) \left(1 + \frac{\varepsilon \frac{M}{m}}{1 - \frac{M}{m}}\right) \frac{1}{(1-\varepsilon)^2} \notag\\
&\leq& \frac{m}{M}\left(1-\frac{M}{m}\right) \left(1 + \frac{\frac{\varepsilon}{1+\varepsilon}}{1 - \frac{1}{1+\varepsilon}}\right)\frac{1}{(1-\varepsilon)^2} \notag\\
&\leq& \frac{m}{M}\left(1-\frac{M}{m}\right) \frac{2}{(1-\varepsilon)^2}.
\end{eqnarray}
Thus, by using (\ref{eq: gap 41}), we obtain
\be
\frac{R(M)}{R^*(M)} \leq \frac{2}{(1-\varepsilon)^2} \times \left\{\begin{array}{cc}4, &  t = \omega(1), \frac{1}{2} \leq M = o(m) \\
6, & M = \Theta(m) \\
\frac{2}{M}, & n = \omega(m), M <\frac{1}{2} 
\end{array}\right.,
\ee

\subsection{Case $t=\Theta(1)$}

In this case, let $n,m \rightarrow \infty$, we have $\frac{M}{m} \rightarrow 0$. 
By using Theorem \ref{theorem: decentralized 1}, we can obtain
\be
R(M) \leq n-t \leq n.
\ee
Then by using (\ref{eq: gap sl 3}), we can obtain
\be
\frac{R(M)}{R^*(M)} \leq 4t. 
\ee  
By using (\ref{eq: achievable dc 1}), we have
\begin{eqnarray}
R(M) &\leq& \frac{m}{M}\left(1-\frac{M}{m}\right) \frac{1}{\rho^2}  \left(1 + f_\rho(t)\right).
\end{eqnarray}
By using (\ref{eq: gap 41}), we obtain
\be
\frac{R(M)}{R^*(M)} \leq \frac{1}{\rho^2}  \left(1 + f_ \rho(t)\right) \left\{\begin{array}{cc}\frac{4t}{\lfloor t \rfloor}, &  n = O(m), t=\Theta(1), \frac{1}{2} \leq M = o(m) \\
\frac{t}{\lfloor t \rfloor}\frac{2}{M}, & n = O(m), n > m, M <\frac{1}{2} \\
2, & n = O(m), n \leq m, M <\frac{1}{2}
\end{array}\right.,
\ee
Denote 
\be
f_g(t) \eqdef \frac{1}{\rho^2}  \left(1 + f_ \rho(t)\right) \left\{\begin{array}{cc}\frac{4t}{\lfloor t \rfloor}, &  n = O(m), t=\Theta(1), \frac{1}{2} \leq M = o(m) \\
\frac{t}{\lfloor t \rfloor}\frac{2}{M}, & n = O(m), n > m, M <\frac{1}{2} \\
2, & n = O(m), n \leq m, M <\frac{1}{2}
\end{array}\right..
\ee
Then, we have
\be
\frac{R(M)}{R^*(M)} \leq \min\{4t, f_g(t)\}.
\ee
Since $\rho = (1-\varepsilon)\rho^*$, where $\varepsilon$ is an arbitrary small positive number and $\rho^*$ is given by Theorem \ref{theorem: caching}, then we obtain that if $t \rightarrow \infty$, $\rho \rightarrow 1-\varepsilon$ and if $t \neq 1$ and $t$ is finite, then $\rho $ is finite, therefore, we can conclude that if $t \neq 1$, then $\frac{1}{\rho^2}  \left(1 + f_ \rho(t)\right)$ is finite.  

\bibliographystyle{IEEEbib}
\bibliography{references}

\end{document}